\documentstyle{mn}

\def\simless{\mathbin{\lower 2pt\hbox
   {$\rlap{\raise 5pt\hbox{$\char'074$}}\mathchar"7218$}}}   
\def\simgreat{\mathbin{\lower 2pt\hbox
   {$\rlap{\raise 5pt\hbox{$\char'076$}}\mathchar"7218$}}}   
\def\be {\begin{equation}}
\def\ee {\end{equation}}

\input psfig

\pubyear{1998}

\title[Mean surface density of companions]{Interpreting the mean surface density of companions in star-forming regions }

\author[M. R. Bate et al.]
  {Matthew R. Bate$^{1}$, Cathie J. Clarke$^2$, and Mark J. McCaughrean$^3$\\
  $^1$ Max-Planck-Institut f\"ur Astronomie, K\"onig\-stuhl 17, D-69117 Heidelberg, Germany\\
  $^2$ Institute of Astronomy, Madingley Road, Cambridge CB3 0HA\\
  $^3$ Max-Planck-Institut f\"ur Radioastronomie, Auf dem H\"ugel 69, D-53121 Bonn, Germany}

\begin{document}
\label{firstpage}
\maketitle

\begin{abstract}

We study the interpretation of the mean surface density of stellar companions 
as a function of separation (or, equivalently, the two point correlation 
function of stars) in star-forming regions.  
First, we consider the form of the functions for various simple stellar 
distributions (binaries, global density profiles, clusters, and fractals) 
and the effects of survey boundaries.  

Following this, we study the dependencies of the 
separation at which a transition from the binary to the large-scale 
clustering regime occurs.  Larson \shortcite{Larson95} found that
the mean surface density of companions follows different power-law 
functions of separation in the two regimes.  
He identified the transition separation with the typical Jeans length
in the molecular cloud.  However, we show that this is valid only for
special cases.  In general, the transition separation depends on 
the volume density of stars, the depth of the star-forming region, the
volume-filling nature of the stellar distribution,
and on the parameters of the binaries.  Furthermore, the
transition separation evolves with time.  We also note that in
young star-forming
regions, binaries with separations greater than the transition separation 
may exist, while in older unbound clusters which have expanded significantly,
the transition contains a record of the stellar density when the stars 
formed.

We then apply these results to the Taurus-Auriga, Ophiuchus, and Orion 
Trapezium star-forming regions.  We find that while the transition 
separation in the Taurus-Auriga star-forming region may indicate a 
typical Jeans length, this is not true of the Orion Trapezium Cluster. 
We caution against over-interpreting 
the mean surface density of stellar companions; while Larson showed that 
Taurus-Auriga is consistent with the stars having a fractal large-scale 
distribution
we show that Taurus-Auriga is also consistent with stars being grouped in
non-hierarchical clusters.  We also argue that to make a meaningful study
of the stellar distribution in a star-forming region requires a 
relatively complete stellar survey over a large area.  
Such a survey does not currently exist for Ophiuchus.
Finally, we show that there is no evidence for sub-clustering or 
fractal structure in the stars of the Orion Trapezium Cluster.  This is
consistent with the fact that, if such structure were present when the
stars formed, it would have been erased by the current age of the cluster
due to the stellar velocity dispersion.

\end{abstract}


\section{Introduction}
\label{introduction}

Stars generally do not form in isolation.  Instead, on small scales, they 
frequently form as members of bound binary or higher-order multiple systems
(e.g. Duquennoy \& Mayor 1991; Mayor et al. 1992; Fischer \& Marcy 1992; 
Ghez, Neugebauer, \& Matthews 1993; Leinert et al. 1993, Simon et al. 1995), 
while on larger scales they are often members of associations 
or clusters of stars (e.g. Gomez et al. 1993; Lada, Strom, \& Myers 1993;
Zinnecker, McCaughrean, \& Wilking 1993).  
Studying the clustering properties of stars on
different length scales may help to determine what processes are involved
in their formation.

Gomez et al. \shortcite{GHKH93} found that the pre-main-sequence stars 
in the Taurus-Auriga molecular cloud are not randomly distributed, 
but instead are in small associations of $\sim 15$ stellar systems
within radii of $\sim 0.5-1.1$ pc.  As one method of analysing the 
spatial distribution of stars, Gomez et al. determined
the two-point angular correlation function and found that it could be 
represented by a single power-law over separations from $0.005$
to $5$ pc, implying that stars are clustered self-similarly.  
However, they also found weak evidence that two-point angular correlation 
function may be better represented by two different power laws with a break
at $\approx 0.05$ pc.

Using data from searches for binary companions to pre-main-sequence stars
in the Taurus-Auriga molecular cloud, Larson \shortcite{Larson95} extended
the two-point angular correlation function to smaller separations than
Gomez et al. \shortcite{GHKH93} and demonstrated that, indeed,
there is a break at $\approx 0.04$ pc.  Rather than using the 
standard two-point angular correlation function, Larson used the
closely-related mean surface density of companions (MSDC) 
(see Section \ref{MCSD}).  The MSDC has the advantage that no 
normalisation is required, whereas the two point correlation function
must be normalised by the average density in the survey area which
can be difficult to determine if the stars are clustered.

Larson \shortcite{Larson95} found that, for stars in the Taurus-Auriga 
molecular cloud, the MSDC has a power-law slope of $\approx -0.6$ on large
scales, but steepens below $\approx 0.04$ pc with a slope of $\approx -2$
on small scales.  The fact that a break occurs 
indicates that a single scale-free process is not responsible for 
the formation of stars on both scales.  The power-law slope of 
$\approx -0.6$ on large scales is due to the clustering of stellar systems
that Gomez et al. \shortcite{GHKH93} studied.  Furthermore, Larson 
pointed out that a power-law slope of $-0.6$ means that the 
number of stars within an angular distance $\theta$ of an average star
increases as $\theta^{1.4}$ and, thus, the distribution of stars on this
scale can be described as a fractal point distribution with dimension 1.4.
Larson identified the power-law slope of $-2$ for small angular separations
with the distribution of binary separations, since stellar pairs closer
than $0.04$ pc in Taurus-Auriga are typically mutually bound.
However, the power-law slope of $\approx -2$ is not due to a
fractal distribution.  Rather, it results from the fact that the frequency
distribution
of binary separations is roughly uniform in log-separation \cite{DuqMay91}.
Finally, Larson noted that the length scale of $\approx 0.04$ pc is 
essentially equal to the typical Jeans length in the Taurus-Auriga
molecular cloud.  Thus, Larson associated the location of the 
break in the MSDC with the Jeans length, speculating that
companions with separations smaller than this formed due to the
fragmentation of a single collapsing molecular cloud core, while on larger
scales stars are grouped self-similarly due to hierarchical structure
in the progenitor molecular clouds.

Following Larson's analysis of the Taurus-Auriga star-forming region (SFR), 
Simon \shortcite{Simon97} considered the spatial distribution of stars
in the Ophiuchus and Orion Trapezium regions.  
As with Taurus-Auriga, a break was found in the MSDC for each region.  
On small scales,
both Ophiuchus and the Orion Trapezium could be fit by power laws with slopes
of $\approx -2$.  On large scales, flatter power
laws were required of $-0.5\pm0.2$ for Ophiuchus and $-0.2\pm0.2$ for 
the Orion Trapezium.  However, the break between the two regimes was found to
occur at $\approx 400$ AU for the Orion Trapezium and
$\approx 5000$ AU for Ophiuchus, compared to $\approx 10000$ AU (taking the
mean of Simon's and Larson's results) for 
Taurus-Auriga.  Simon concluded that all three SFRs 
had similar distributions
of binary separations and similar fractal structure on large scales, but
that the location of the break seemed to depend not only on the Jeans length,
but also on the stellar density of the SFR.

Finally, Nakajima et al. \shortcite{NTHN98} considered the MSDC
of stars in the Orion, Ophiuchus, Chamaeleon,
Vela, and Lupus star-forming regions.  Again, for those regions where the
survey data extends to small enough separations, they find a break in
the MSDC with a power-law slope of $\approx -2$ on small scales and 
power-law slopes ranging from $-0.15$ to $-0.82$ on large scales.  The location
of the break was also found to vary from a minimum of $\approx 1000$ AU to
a maximum of $\approx 30000$ AU.  Nakajima et al. also considered the
nearest-neighbour distributions for each of the regions and found that
when the nearest-neighbour distribution could be fit well by a Poisson
distribution, the MSDC had a power-law index close to zero on large-scales,
while when the nearest-neighbour distribution was broader than the Poisson
distribution, the MSDC had a large, negative power-law index.
They interpreted this as evidence that the MSDC may indicate a star
formation history in the region rather than the presence of self-similar
spatial structure; if the stars have a range of ages, the older stars 
typically will be more dispersed than the younger stars resulting a spread
in the distribution of separations of nearest neighbours and a range of
stellar surface density which provides the slope of the large-scale MSDC.

Motivated by these papers, we make a careful study of the 
interpretation of the mean surface density of companions (MSDC) 
of star-forming regions.  Amongst other goals, we wish to determine the
relationship of the break between the binary and large-scale regimes to the
Jeans length and the stellar density in star-forming regions.  We also
want to determine how sensitive the MSDC is to detecting sub-structure in
a stellar distribution and, when detected, what can be said about the form
of the sub-structure (e.g. whether the sub-structure is self-similar or not) 
and how robust the result is.

In Section \ref{MCSD} we consider the calculation of the MSDC function,
handling of survey boundaries, and the results for simple stellar
distributions (binaries, global density profiles, clusters, and fractals).  
In Section \ref{posbreak} we derive the dependencies of the break between
the binary and large-scale regimes, and show that the separation at which 
the break occurs can only be identified with the Jeans length in special
cases.  We also indicate how the MSDC of SFRs is expected to evolve with time.
Based on these results, we reconsider the Taurus-Auriga, Ophiuchus, and
Orion Trapezium star-forming regions in Section \ref{application}.
Finally, we present our conclusions in Section \ref{conclusions}.

\section{The mean surface density of companions}
\label{MCSD}

In the simplest case, the mean surface density of companions
can be determined as follows.
For each star, calculate the angular separation $\theta$ to all other stars.
Bin the separations resulting from these stellar pairs into annuli of 
separation $\theta$ to $\theta + \delta\theta$.  The binning is most
conveniently done using annuli with logarithmically increasing radii.  
The number of separations from such pairs, $N_{\rm p}$,
within each size of annulus is then divided by the area of the annulus and 
averaged by dividing by the total number of stars $N_*$ giving the mean 
surface density of companions as a function of separation, 
$\Sigma_{\rm com}(\theta) = N_{\rm p}/(2\pi~\theta~\delta \theta~N_*)$.  
We will refer to this as Method 1.
The MSDC is related to the angular two-point
correlation function, $w(\theta)$,
by $\Sigma_{\rm com}(\theta) = (N_*/A)(1+w(\theta))$,
where $A$ is the survey area
\cite{Peebles80}.

\subsection{Handling boundaries}
\label{boundaries}

For a survey region of finite size, using the above method (Method 1), 
to calculate $\Sigma_{\rm com}(\theta)$, results in some of the 
companions to stars closer than $\theta$ to a boundary being missed.
This has little effect if the survey region is 
large in comparison to $\theta$, since only a small fraction of the
stars are affected.  However, when $\theta$ becomes large, the missing
companions result in an unphysical drop in $\Sigma_{\rm com}(\theta)$.

Several methods for avoiding such boundary effects have been used 
(see Peebles 1980, and references within).
One method is to select an inner subsample of
$N_1$ stars such that none is closer than $\theta_{\rm max}$ from a boundary 
and for each of them calculate the number of companions from the full
sample.  $\Sigma_{\rm com}(\theta)$ is then normalised by dividing by $N_1$
rather than $N_*$.  This method (Method 2) avoids boundary effects, 
but also eliminates some information and does not probe the full range of
separations.

A modification of Method 2 (Method 3) is to select a different inner
subsample for each size of annulus, $\theta_{\rm i}$ to 
$\theta_{\rm i} + \delta\theta$,
such that each subsample of $N_{\rm i}$ stars are all
farther than $\theta_{\rm i} + \delta\theta$ from a boundary.  
This allows the maximum information to be used for each separation, but again
the full range of separations cannot be studied since as $\theta_{\rm i}$
increases, $N_{\rm i}$ decreases.

Other methods try to correct for the presence of boundaries.  In Method 4,
the area of each annulus is directly computed for each star (subtracting that
part which falls outside a boundary) to give a corrected surface density 
of companions for that star before the mean over all stars is calculated.
Rather than calculate the area of each annulus explicitly (which can be
difficult), a similar method (Method 5) calculates a 
Monte Carlo estimate of the area of each annulus that lies within the 
survey boundaries.  This is achieved as follows:
$N_{\rm t}$ points are placed at random in the survey area, $A$;
$\Sigma_{\rm com}(\theta)$ is calculated using Method 1 for both the 
trial points 
($\Sigma_{\rm trial}$) and for the real data ($\Sigma_{\rm data}$); then the
corrected $\Sigma_{\rm com}(\theta)$ of the real data is given by
\be
\Sigma_{\rm corr}(\theta) = \frac{\Sigma_{\rm data}(\theta)}{\Sigma_{\rm trial}(\theta)} \frac{N_{\rm t}}{A}.
\ee
Methods 4 and 5 correct for boundaries if the distribution of
sources on scales of order the survey size is uniform (hence their use in 
studying the large-scale structure of the universe).  However,
if there is a surface density gradient across the boundary
these methods {\em do not} fully correct for boundaries, as
demonstrated in Section \ref{global}.

\subsection{Binaries}
\label{binaries}

Larson \shortcite{Larson95} found that the binary systems in Taurus-Auriga
are characterised by an MSDC with a power-law slope of $\approx -2$, 
which can be easily shown to be equivalent to the distribution of binary 
separations being uniform in the logarithm of separation.
Take $N$ binary systems with separations, $r$, distributed as 
\be
\label{eqsepdist}
\frac{{\rm d}N}{{\rm d}({\rm log}(r))} = k,
\ee
where $k$ is a constant.  Then
\be
\label{dNequation}
{\rm d}N = \frac{k}{r} {\rm d}r,
\ee
is the number of companions with separations in the range 
$r$ to $r + {\rm d}r$.  Taking a two-dimensional projection, the range 
$r$ to $r + {\rm d}r$ defines an annulus of area $2\pi r~{\rm d}r$.
Therefore, the MSDC is
\be
\Sigma_{\rm com}(r) = \frac{1}{N} \frac{{\rm d}N}{2\pi r~{\rm d}r} = \frac{1}{N} \frac{k}{2\pi r^2},
\ee
which has a power-law slope of $-2$.  Integrating equation \ref{dNequation}
to find $k$ gives
\be
\Sigma_{\rm com}(r) = \frac{1}{2\pi~{\rm ln}[R_{\rm max}/R_{\rm min}]~r^2},
\ee
where $R_{\rm max}$ and $R_{\rm min}$ are the maximum and minimum binary
separations, respectively.  Finally, if not all systems are binaries
\be
\label{eqbinaryMSDC}
\Sigma_{\rm com}(r) = \frac{B_{\rm freq}}{2\pi~{\rm ln}[R_{\rm max}/R_{\rm min}]~r^2},
\ee
where $B_{\rm freq}$ is the binary frequency 
(defined as the number of binary systems
over the total number of systems).

Thus, Larson's \shortcite{Larson95} result for the Taurus-Auriga 
pre-main-sequence binaries is in general agreement with the period distribution
for the main-sequence G-dwarf primaries of 
Duquennoy \& Mayor \shortcite{DuqMay91} which, 
in the logarithm of period, is flat to first order.

Finally, we note that fitting power-law slopes to the MSDC and comparing
the values of the indices is not the best
way to compare the distribution of binary separations between star-forming
regions since it ignores any deviations from a power-law.  Instead, it is 
preferable to take the usual approach, namely comparing
the fractional number of binaries (the number of binaries in a range of 
separations over the total number of systems) as a function of the logarithm
of separation.

\subsection{Global density profiles}
\label{global}

The MSDC for stars distributed uniformly is independent of separation and 
simply equal to the mean surface density of stars over the whole survey 
region.  If the stars are clustered 
(Section \ref{clusters}) or distributed in a non-uniform, self-similar 
(fractal) distribution, the MSDC is a function 
of separation (e.g. Larson 1995).   However, a particular MSDC
does not correspond to a unique two-dimensional
distribution of stars.  In particular, an MSDC
which has a power-law dependence on separation does not necessarily correspond
to a self-similar or fractal distribution of stars; a simpler
surface density distribution (e.g. a centrally-condensed
density profile) can give the same slope over a large separation range.  

In this section, we consider the MSDC of global density profiles.  
We also study the effects that the different ways
of handling boundaries (Section \ref{boundaries}) have on the MSDC
when the underlying density profile is not uniform.

\begin{figure}
\vspace{0.2truecm}
\vspace{13.5truecm}
\caption{\label{global1} The mean surface density of companions $\Sigma_{\rm com}(r)$ for four different distributions of $10^4$ single stars (shown above).  The distributions have stellar surface densities $\Sigma(x) \propto x^{\gamma}$ with $\gamma=0$ (top left, solid line); $\gamma=1$ (top right, dotted line); $\gamma=2$ (lower left, dashed line); and $\gamma=5$ (lower right, long-dashed line).  To avoid boundary effects the distributions are continued outside the illustrated area.  In each case, $\Sigma_{\rm com}(r)$ and its dispersion are calculated from 20--100 random renderings of the distributions.}
\end{figure}

\begin{figure}
\vspace{0.2truecm}
\centerline{\hspace{0.2truecm}\psfig{figure=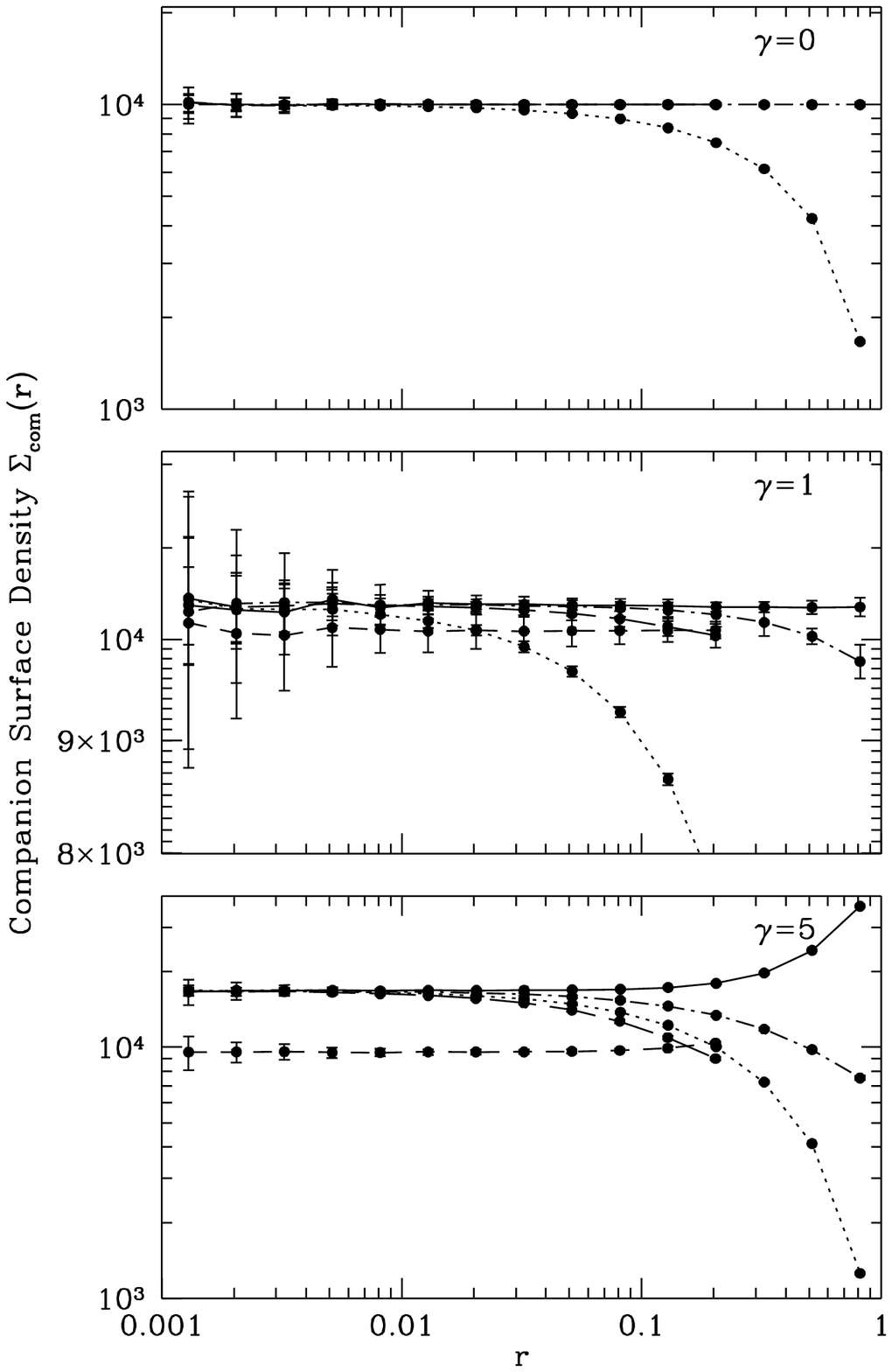,width=13.0truecm,height=13.0truecm,rwidth=9.0truecm,rheight=13.5truecm}}
\caption{\label{global2} The mean surface density of companions $\Sigma_{\rm com}(r)$ for the distributions from Figure \ref{global1} with $\gamma=0$, 1, and 5, but comparing the results from Methods 1 (dotted lines), 2 (dashed lines), 3 (long-dashed lines), and 5 (dot-dashed lines) to the MSDCs without boundary effects from Figure \ref{global1} (solid lines). }
\end{figure}

\subsubsection{Density gradients}
\label{globaldensitygradient}

Consider a stellar distribution with a gradient in surface density 
along one direction $\Sigma(x) \propto x^{\gamma}$ with $x>0$.  Now, consider
the surface density of companions in an annulus of radius $r_{\rm ann}$ 
and width $\delta r$ centred on one star in the distribution with an 
$x$-coordinate of $x_*$.  
To first order, the surface density at $x_*+\delta x$ 
differs from the surface density at $x_*$ by $1+\gamma~\delta x/x_*$.  Thus,
as long as $r_{\rm ann} \ll x_*$, the mean surface density around the 
annulus, $\Sigma_{\rm ann}(x_*)$,
is approximately equal to the surface density at $x_*$, $\Sigma(x_*)$, 
since the
first-order term cancels and only leaves second-order and higher effects:
for $\gamma=0$ or $\gamma=1$ the equality is exact.  
If each star in the distribution satisfies $r_{\rm ann} \ll x_*$, 
the MSDC should be almost independent of separation.  
For $\gamma=0$ the value is simply
the mean density in the survey area.  However, for $\gamma \neq 0$ the value 
is greater since the surface density distribution is sampled
more in the high-density regions and less in the low-density regions 
(the sampling points are at the stellar positions) 
and, hence, a {\em surface-density-weighted} mean surface density is obtained.

The MSDC for distributions with $\gamma=0, 1, 2,$ and 5 in
Figure \ref{global1} demonstrate the lack of dependence of the MSDC 
on separation for such density gradients 
and the higher overall values for $\gamma \neq 0$.  The increase at large
radii for $\gamma=2$ and $\gamma=5$ is due to the second-order 
and higher effects which become important 
when $r_{\rm ann}$ is of order the survey area's dimensions.
Note that in these figures, and all similar 
ones in this paper, the errorbars are determined by randomly producing 
20-100 distributions of stars that follow the prescribed surface density 
distribution and calculating the mean and standard deviation of the MSDC at 
each separation.

\begin{figure}
\vspace{0.2truecm}
\vspace{8.0truecm}
\caption{\label{globaldiag} Diagram showing the sampling regions of small and large annuli centred on a star a distance $r_*$ (solid line) from the centre of the global density maximum of a stellar distribution with the radial surface density profile $\Sigma_{\rm r} \propto r^{-1}$.  The annuli of radii $r_{\rm ann}$ (dashed lines) have width $\delta r$.  Notice that the mean surface density around the small annulus is approximately equal to the surface density at $r_*$ ($\Sigma_{\rm ann}(r_*) \approx \Sigma_{\rm r}(r_*)$), while the surface density around the large annulus is always less than that at $r_*$ and, thus, so is the mean ($\Sigma_{\rm ann}(r_*) < \Sigma_{\rm r}(r_*)$).  In the limit that $r_{\rm ann} \gg r_*$, $\Sigma_{\rm ann}(r_*)$ tends towards $\Sigma_{\rm r}(r_{\rm ann})$. }
\end{figure}

The MSDC functions in Figure \ref{global1} were produced using Method 1 in 
Section \ref{boundaries}, but the stellar distributions were continued
outside of the regions shown in Figure \ref{global1} to avoid missing 
neighbours at the edges.  However, for a real stellar survey, there is 
no knowledge of the stellar distribution outside of the survey boundaries.  
Therefore, in Figure \ref{global2}, we give the MSDC using only the 10000 
stars in the area $1 \leq x \leq 2$ and $0 \leq y \leq 1$, calculated using
Methods 1, 2, 3, and 5.  Since Methods 2 and 3 only use subsets of the data
they are only calculated to separations of 1/5 of the survey area's size as
this gives a good trade-off between the range of separations that are
covered and the number of stars used for the calculation.
Method 4 is not used here since it gives
a similar result to Method 5, but is much more difficult to calculate 
for a simple survey area let alone for an irregular one, since the 
area of the part of each annulus that falls within the
boundaries has to be determined.  Throughout this paper,
Method 5 is calculated using 10000 trial points in the survey area.

For the uniform density case ($\gamma=0$), all methods, except
Method 1, return the `correct' MSDC (i.e. that given by extending the
stellar distribution outside the boundaries of the survey region).  
Method 1, in which there is no attempt
to correct for the effect of the boundaries, gives an MSDC which differs 
appreciably from the correct MSDC even on scales less than a order of
magnitude smaller than the size of the survey region.  In particular,
if a power-law slope is fit to the MSDC even between such small separations 
as 1/100 to 1/10 of the survey's dimensions, a slope of $-0.055 \pm 0.003$ 
is derived, rather than a slope of zero.

Due to the density gradients across the boundaries, none of the methods 
give the `correct' MSDC for the whole stellar distribution when 
$\gamma \neq 0$.  As for $\gamma=0$, using Method 1 results in a rapid
drop in the MSDC as the separation increases.  Method 2 gives the correct
MSDC, but over a smaller region than the whole survey area.  
With $\gamma \neq 0$, this results in a different normalisation because the
mean surface density over the smaller area is different to that over the
whole survey area.  Method 3 gives the correct normalisation on small-scales,
but still results in a rapid fall-off at large scales which may even exceed
that of Method 1 (e.g. in the case where $\gamma=5$).  
Method 5 gives the correct normalisation on small-scales
and partially corrects for the fall-off at large separations.  

Finally, although none of the methods correct perfectly for the survey
boundaries when $\gamma \neq 0$, all except Method 1 give a 
good approximation (i.e. an error of $\simless 0.1$ dex) 
to the `correct' MSDC if $0 \leq \gamma \simless 2$.

\subsubsection{Centrally-condensed stellar distributions}
\label{globalcondensed}

Next, we consider a system of stars distributed according to the radial surface
density distribution $\Sigma_{\rm r}(r) \propto r^{-\alpha}$ with 
$\alpha \geq 0$.
Consider the mean surface density of stars in an annulus of radius
$r_{\rm ann}$ centred on one star in this distribution with distance 
$r_*$ from the position of peak surface density (Figure \ref{globaldiag}).  
If $r_{\rm ann} \ll r_*$, the mean surface density of stars in the annulus,
$\Sigma_{\rm ann}(r_*)$, is 
approximately equal to the local surface density $\Sigma_{\rm r}(r_*)$.
On the other hand, if $r_{\rm ann} \gg r_*$, the mean surface 
density of stars in the annulus is less than $\Sigma_{\rm r}(r_*)$ 
and tends towards $\Sigma_{\rm r}(r_{\rm ann})$ 
(see Figure \ref{globaldiag}).

\begin{figure}
\vspace{0.2truecm}
\vspace{13.5truecm}
\caption{\label{global3} The mean surface density of companions $\Sigma_{\rm com}(r)$ for four different distributions of $10^4$ single stars (shown above).  The four distributions have radial surface density profiles given by: a) $\Sigma_{\rm r} =$ constant (top left, solid line), b)  $\Sigma_{\rm r} \propto r^{-1/2}$ (top right, dotted line), c)  $\Sigma_{\rm r} \propto r^{-1}$ (lower left, short-dashed line), d)  $\Sigma_{\rm r} \propto r^{-3/2}$ (lower right, long-dashed line).  Global surface density profiles less extreme than $\Sigma_{\rm r} \propto r^{-1}$ result in essentially flat MSDC functions.}
\end{figure}

\begin{figure}
\vspace{0.2truecm}
\centerline{\hspace{0.2truecm}\psfig{figure=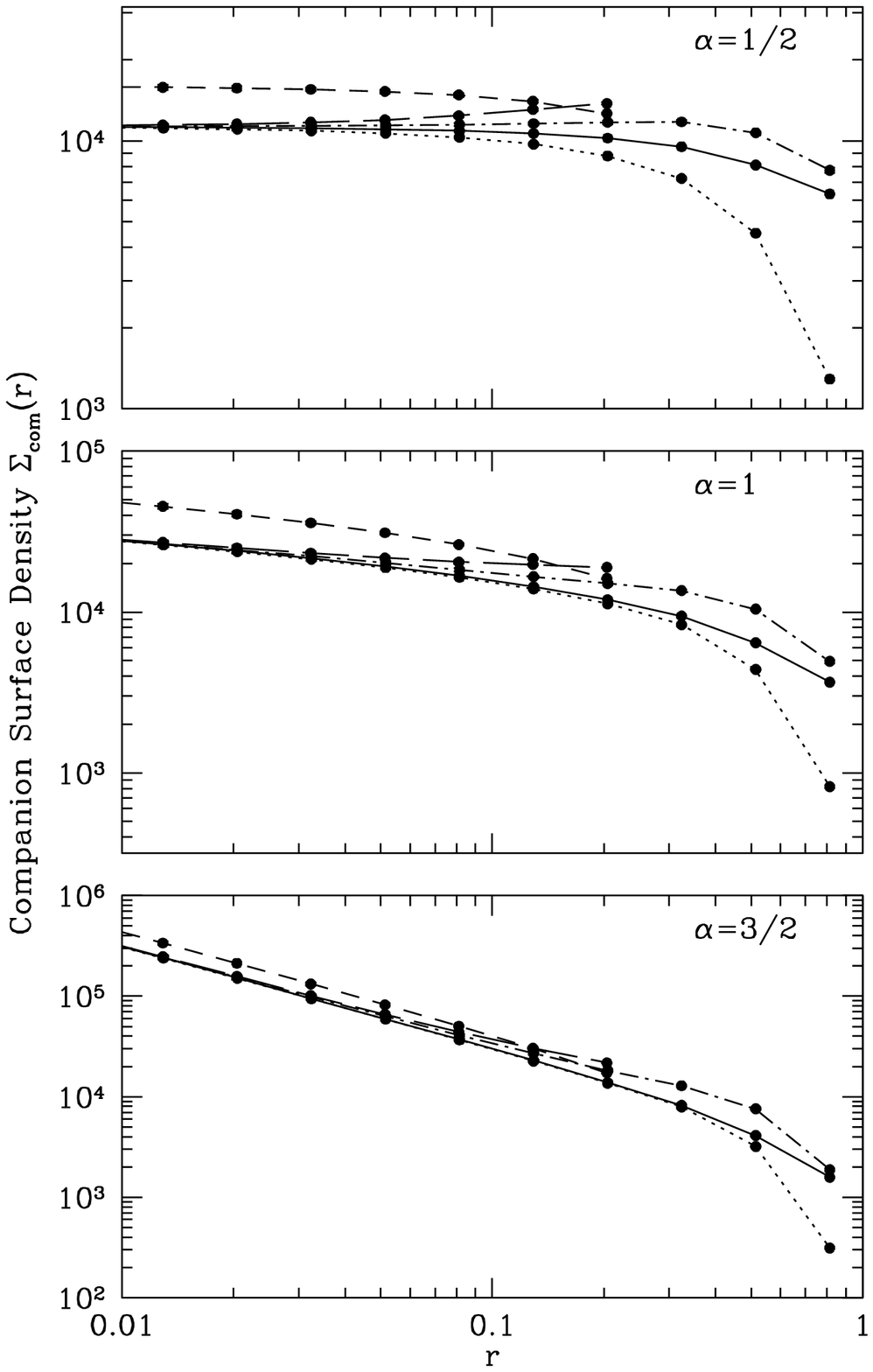,width=13.0truecm,height=13.0truecm,rwidth=9.0truecm,rheight=13.5truecm}}
\caption{\label{global4} The mean surface density of companions $\Sigma_{\rm com}(r)$ for the distributions from Figure \ref{global3} with $\alpha=1/2$, 1, and $3/2$, but comparing the results from Methods 1 (dotted lines), 2 (dashed lines), 3 (long-dashed lines), and 5 (dot-dashed lines) to the MSDCs without boundary effects from Figure \ref{global3} (solid lines).  In none of the cases do methods 1-5 give perfect correction for boundary effects. }
\end{figure}

Now, consider the {\em mean} surface density of companions over {\em all} 
stars $\Sigma_{\rm com}(r_{\rm ann})$ 
in a survey region that is centred on the peak surface density.
The variation of the MSDC as a function of separation depends on the fraction 
of the stars that have $r_{\rm ann} \simgreat r_*$.  This fraction
increases as $r_{\rm ann}$ is increased, but it also depends on $\alpha$.
For example, for $\alpha=1$, the number of stars within $r_{\rm ann}$ 
of the central peak 
(i.e. with $r_*<r_{\rm ann}$) increases linearly with $r_{\rm ann}$.
Thus, for larger annuli, more and more stars have
$\Sigma_{\rm ann}(r_*) < \Sigma_{\rm r}(r_*)$ but, only when
$r_{\rm ann}$ is greater than half the radius of the survey area,
$R_{\rm sur}$, do less than 50\% of the stars have
$\Sigma_{\rm ann}(r_*) \approx \Sigma_{\rm r}(r_*)$.  
When $r_{\rm ann} > R_{\rm sur}/2$, most of the stars tend towards
$\Sigma_{\rm ann}(r_*) < \Sigma_{\rm r}(r_{\rm ann})$.
Thus, the MSDC should only decrease slowly as the separation increases, 
due to the slowly changing number of stars with $r_*<r_{\rm ann}$ until
$r_{\rm ann}$ approaches $R_{\rm sur}$ at which stage there is a transition
from the slowly decreasing slope to power-law slope of $-\alpha$.

For $\alpha<1$, the number of stars that have $r_*<r_{\rm ann}$ 
increases even more slowly with increasing $r_{\rm ann}$.  Also, since the
radial surface density varies less overall, $\Sigma_{\rm ann}(r_*)$ is
closer to $\Sigma_{\rm r}(r_*)$ even when $r_*<r_{\rm ann}$.  Thus, the
MSDC depends less on separation.  The
extreme case is for $\alpha=0$ when, since $\Sigma_{\rm r}(r_*)$ is constant,
$\Sigma_{\rm ann}(r_*)$ is always equal to $\Sigma_{\rm r}(r_*)$ and so
the MSDC is constant.

For $\alpha>1$, however, the number of stars with $r_*<r_{\rm ann}$
increases quicker with increasing $r_{\rm ann}$ and the overall variation of
the radial surface density is more extreme, thus, the MSDC,
$\Sigma_{\rm com}(r_{\rm ann})$, falls faster with increasing $r_{\rm ann}$.
In the limit that $\alpha \rightarrow \infty$, all the
stars are in the central peak and for all stars 
$\Sigma_{\rm ann}(r_*) = 
\Sigma_{\rm r}(r_{\rm ann}) \propto {r_{\rm ann}}^{-\alpha}$ so that 
the MSDC has a power-law slope of ${-\alpha}$ for all separations
($\Sigma_{\rm com}(r_{\rm ann}) \propto {r_{\rm ann}}^{-\alpha}$).

In summary, the MSDC of a uniform distribution of stars is independent of 
separation, while in the limit $\alpha \rightarrow \infty$, the MSDC 
has a power-law of slope $-\alpha$.  For intermediate cases, the slope over
most separations is less than $-\alpha$ with a transition towards a slope of
$-\alpha$ when $r_{\rm ann} \approx R_{\rm sur}$.

In Figure \ref{global3}, we give the MSDC for global density profiles
$\Sigma_{\rm r}(r) \propto r^{-\alpha}$ where $\alpha=0$, $1/2$, 1, and $3/2$.
For separations less than 1/10 of the dimension of the survey region the 
MSDC have slopes of $0$, $-0.013 \pm 0.004$, $-0.22 \pm 0.01$, 
and $-1.01 \pm 0.01$, respectively.  This is in agreement with the above
analysis (i.e. the slopes are shallower than $-\alpha$ if $\alpha>0$).
It is important to notice that {\em the MSDC can
have a power-law slope even without any
sub-clustering or hierarchical structure} (e.g.~see the Jones \& Walker MSDC
for the Orion Trapezium Cluster 
in Section \ref{otc} and Figure \ref{joneswalker}).  As predicted above, 
the slope deviates from a power-law only for separations 
$\simgreat 1/10$ the survey region size (i.e. when 
$r_{\rm ann} \approx R_{\rm sur}$), when there is a transition towards a
slope of $-\alpha$.  The maximum difference between the small-scale 
and large-scale power-law slopes occurs for $\alpha \approx 1$.

\begin{figure}
\vspace{0.2truecm}
\vspace{8.0truecm}
\caption{\label{cluster1} The mean surface density of companions $\Sigma_{\rm com}(r)$ for three different distributions of clusters of stars (shown above).  Each cluster consists of 10 stars distributed uniformly within a sphere of radius 0.025.  The three distributions consist of: a) 10 clusters (left, solid line), b) 100 clusters (centre, dotted line), c) 1000 clusters (right, dashed line).  The clusters are randomly distributed.  Cases b) and c) can be thought of as having the same volume density of clusters as case a), but being 10 and 100 times deeper, respectively.  The higher the mean density of stars, the more difficult it is to detect the clustering.}
\end{figure}

Although all centrally-condensed global density profiles show 
power-law slopes in the derived MSDC, for $\alpha \simless 1$ the MSDC
is rather flat, depending only weakly on separation.  This is important
because most young clusters are likely to be less centrally condensed
than the $\alpha=1$ case (e.g McCaughrean \& Stauffer 1994), 
which corresponds to a three-dimensional density distribution of 
$\rho(r) \propto r^{-2}$, the singular isothermal sphere.  Therefore,
if a cluster has an overall radial surface density profile and {\em also}
contains sub-clustering and/or hierarchical structure, the sub-structure
will generally be detectable because it will steepen the slope of the
MSDC and/or perturb it from a true power-law (e.g.~Section 
\ref{orionsubcluster}).  Thus,
{\em the MSDC can still be used to look for
sub-clustering and/or hierarchical structure even where
an overall global density distribution exists}.  

\begin{figure}
\vspace{-0.2truecm}
\centerline{\hspace{0.5truecm}\psfig{figure=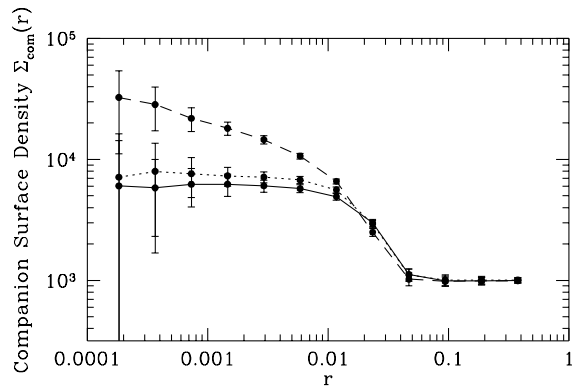,width=12.0truecm,height=12.0truecm,rwidth=9.0truecm,rheight=6.0truecm}}
\caption{\label{cluster2} The mean surface density of companions $\Sigma_{\rm com}(r)$ for three different distributions of clusters of stars.  In each case, the distributions consist of 100 randomly-positioned clusters of 10 stars (i.e. as in case b) in Figure \ref{cluster1}).  The clusters have radii of 0.025 and the distributions of stars within the clusters are given by: a) $\rho =$ constant (solid line), b) $\rho \propto r^{-1}$ (dotted line), and c) $\rho \propto r^{-2}$ (dashed line).}
\end{figure}

\begin{figure}
\vspace{0.2truecm}
\vspace{12.0truecm}
\caption{\label{cluster3} The mean surface density of companions $\Sigma_{\rm com}(r)$ for four different distributions of $10^3$ single stars (shown above).  The four distributions differ in the fraction of stars that are in clusters: a) 100\% (top left, solid line), b) 70\%  (top right, dotted line), c) 30\% (lower left, short-dashed line), d) 10\% (lower right, long-dashed line).  Each cluster consists of 10 stars distributed uniformly within a sphere of radius 0.025.  The lower the fraction of stars that are in clusters, the more difficult the presence of clusters is to detect.}
\end{figure}

As in Section \ref{globaldensitygradient}, a centrally-condensed global 
density profile means that the boundaries of the survey regions are difficult 
to correct for (Figure \ref{global4}).  
Again, Method 1 results in a fall-off for large separations
and Method 2 gives a different normalisation (as well as some information
being lost).  Method 3 over-corrects for the fall-off at
large separations.  Both Methods 4 and 5 assume that the surface density of
companions around an annulus is constant.  With a falling global density
profile, these methods
result in an over-estimate of the mean surface density
of an annulus, since the low-density region of the annulus is missing.
This results in an over-correction for the MSDC.
Finally, note that in calculating the slope of the MSDC for separations 
$\simgreat 1/10$ of the survey's smallest dimension, the relative error
decreases as $\alpha$ increases (c.f. $\alpha=1/2$ and $\alpha=3/2$).

\subsection{Clusters}
\label{clusters}

Gomez et al. \shortcite{GHKH93} found that stars in the Taurus-Auriga SFR
are not distributed randomly, but instead are clustered into small groups.
It is therefore of interest to consider the MSDC of clusters of stars.
Examples of the MSDC derived for randomly-distributed simulated clusters
are given in Figure \ref{cluster1}.
They are characterised by an MSDC equal to the global mean surface 
density on large scales, but have larger than average surface densities
on small scales with a transition between the two regimes for separations 
of order the clusters' radii.

On scales less than the cluster radius, the MSDC
depends on the spatial distribution of stars within the clusters.
Clusters with uniform volume density (Figure \ref{cluster1})
give a flat MSDC on small scales, while centrally-condensed clusters 
give an MSDC that rises as the separation is decreased (Figure \ref{cluster2}).
This is analogous to the power-law slopes of the MSDCs that result from the
centrally-condensed global density profiles in Section \ref{globalcondensed}.
However, the clusters have to be quite
centrally condensed for there to be much effect.  Clusters with
density profiles of $\rho \propto r^{-2}$ have power-law 
slopes of $-0.22 \pm 0.01$
on small scales (as determined in Section \ref{globalcondensed}), but
clusters with $\rho \propto r^{-1}$ still give an essentially flat MSDC.
Therefore, it is difficult to determine whether or not clusters are
centrally condensed unless they have density profiles of $\rho \propto r^{-2}$
or steeper (i.e.~$\Sigma_{\rm r}(r) \propto r^{-1}$).

Given a distribution of stars, it important to be able to detect whether
the stars are clustered.  The
sensitivity of the MSDC to clustering depends on the crowding of the clusters
and on the fraction of stars that are not members of clusters.  Figure
\ref{cluster1} demonstrates that the more crowded clusters are, the more
difficult it becomes to detect them.  Well-separated clusters are 
detectable because the mean surface density of companions on small scales
(within a cluster) is larger than the global mean surface density.  If
the global mean surface density approaches or is greater than 
the mean surface density of companions within a cluster, the confusion 
limit is reached; the clusters overlap and it is no longer
possible to distinguish that the stars are clustered.  The confusion limit 
can be approached either because the clusters actually overlap, or because
a three-dimensional distribution of clusters is projected on to the 
two-dimensional plane of the sky.  The rate at which the confusion limit
is reached with increasing depth depends on the volume-filling factor;
if the clusters are well-separated in three-dimensions, a larger depth
is required before the clusters begin to overlap.  The detection of
binaries (effectively very simple clusters) suffers from the same effect
(see Section \ref{posbreakSFR}).

A similar effect can be obtained by including unclustered stars 
(Figure \ref{cluster3}).  Even if
clusters are well-separated on the sky, stars that are not cluster members
(e.g. randomly-distributed stars) decrease the difference between the
mean surface density of companions on small and large scales.  
Unclustered stars have a low companion surface densities on small scales.  
Thus, the more unclustered stars there are, the lower the {\em mean} 
surface density of companions becomes on small scales and the closer it
becomes to the global mean surface density.  In Figure \ref{cluster3} 
one can easily detect when $30-100\%$ of the stars are in
clusters, but the detection when 10\% of the stars are in 
clusters is
only a $\approx 3\sigma$ result at separations close to the radii of
the clusters with an even a less significant detection at smaller 
separations.

\subsection{Self-similar distributions}
\label{selfsimilar}

\begin{figure*}
\vspace{0.2truecm}
\vspace{8.5truecm}
\caption{\label{fractal1} The mean surface density of companions $\Sigma_{\rm com}(r)$ for three different self-similar (fractal) stellar distributions.  The three distributions have fractal dimensions a) $F_{\rm dim}=1.4$ (left), b) $F_{\rm dim}=1.9$ (centre), c) $F_{\rm dim}=2.4$ (right).  For each case, $\Sigma_{\rm com}(r)$ is given for stellar distributions with depths of 1 (solid line and upper left graph), 10 (dotted line and upper right graph), and 100 (dashed line and lower left graph) times the horizontal or vertical dimension of the survey region.  Examples of the stellar distributions are given above the graphs of $\Sigma_{\rm com}(r)$, with the lower right graphs giving the distribution of stars in the z (depth) dimension.  Notice that for higher $F_{\rm dim}$ the stars become more uniformly distributed with depth.  Also, with a low $F_{\rm dim}$, as the depth is increased the number of stars is increased, but the addition stars typically do not appear in regions already occupied by closer stars.  Thus, the MSDC on small-scales is relatively independent of the depth for low fractal dimensions.}
\end{figure*}

Larson \shortcite{Larson95} found that the MSDC of the stars in the 
Taurus-Auriga SFR could be well fit by a power-law slope of 
$\approx -0.6$ on large
scales ($\simgreat 0.04$ pc).  In general, the number of stars within an
angular distance $\theta$, on the sky, of an average star increases as 
$\theta^{({\rm slope}+2)}$.  For a fractal point distribution with fractal
dimension $F_{\rm dim} \leq 2$, the number of stars closer than 
$\theta$ increases as $\theta^{F_{\rm dim}}$.  
Thus, Larson noted that the large-scale distribution of
stars in the Taurus-Auriga SFR could be described as a fractal 
point distribution with fractal dimension 
$F_{\rm dim} = {\rm slope} + 2 \approx 1.4$.

In this section we consider the MSDC of self-similar (or fractal) stellar
distributions.  Figure \ref{fractal1} gives the MSDCs for stellar
distributions with fractal dimensions of $F_{\rm dim}=1.4$, 1.9, and 2.4.
The distributions were produced using a box-counting algorithm: a cube
of side-length $L$ is divided into $N_{\rm div}^3$ sub-cubes of 
side-length $L/N_{\rm div}$; $N_{\rm ran}$ of these sub-cubes
are randomly selected; the process is repeated recursively, terminating
at the desired level of recursion when each of the smallest sub-cubes 
has a star placed in it.  The fractal dimension is given by $F_{\rm dim} = 
{\rm log}_{\rm e}(N_{\rm ran})/{\rm log}_{\rm e}(N_{\rm div})$.
To obtain a specific number of stars, some of the
stars are randomly selected and removed.  To get a rectangular volume 
that has one dimension longer than the other two, a fractal distribution 
is created for a cube of the largest dimension, and the desired 
rectangular volume is given by a sub-volume of the cube.

For each fractal dimension, we examine the dependence 
of the MSDC on the effects of the depth
of the stellar distributions.  In Section \ref{clusters} and 
Figure \ref{cluster1}, we found that it becomes more difficult to
detect clustering as the clusters become more crowded on the sky.  If clusters
are distributed randomly in three dimensions, then looking through 10 times
the depth, raises the MSDC on large scales by an order of magnitude and 
decreases the difference between the MSDC on small and large scales.

With fractal stellar distributions, the effect of the depth of the SFR
depends on the fractal dimension (or volume-filling nature)
of the stellar distribution.  Some examples are as follows.
A random (or uniform) point distribution in three dimensions has a fractal 
dimension $F_{\rm dim}=3$, since the number of companions within a distance $r$
of a typical star increases as $r^3$.  A uniform two-dimensional
distribution has a fractal dimension $F_{\rm dim}=2$, while a uniform linear 
distribution has $F_{\rm dim}=1$, and a single point has a fractal dimension of $F_{\rm dim}=0$.  Other distributions with 
the same fractal dimensions as those above are possible 
(except for $F_{\rm dim}=0$).
For a uniform three-dimensional stellar distribution ($F_{\rm dim}=3$), 
if the depth is
increased by an order of magnitude, then the MSDC is also increased by an
order of magnitude independent of separation 
(e.g. Figures \ref{cluster1} and \ref{posbreak1}).
However, in general, for a three-dimensional stellar distribution with 
$F_{\rm dim}<3$, the MSDC increases by {\em less} than one order of magnitude 
if the depth is increased by 
one order of magnitude, since the number of companions does not increase 
linearly with depth.  In general, the MSDC is increased (on the largest 
scales) by
\be
\label{depthscaling}
{\rm MSDC} \propto D^{F_{\rm dim}/3},
\ee
where $D$ is the depth.
In the limit that $F_{\rm dim}=0$, there is only ever one object
no matter what factor the depth is increased by.  Note however, that the
dependence of the MSDC on depth
depends somewhat on the specific distribution and/or orientation.  For example,
in the case of a uniform linear distribution, if the depth increases
parallel to the line of objects, then the MSDC will increase linearly
with depth, while if the depth increases perpendicular to the line of
objects, increasing the depth will have no effect on the MSDC.  In general
(i.e. when all possible distributions are averaged over), 
however, the MSDC follows equation \ref{depthscaling} on the largest scales.

The scaling of equation \ref{depthscaling} is apparent in Figure 
\ref{fractal1}; in general, on the largest scales, increasing the depth 
has a greater effect on the MSDC for
distributions with greater fractal dimensions 
(greater volume-filling factor).   However, note that
increasing the depth affects the MSDC on large scales more than
on small scales and this effect is more apparent for lower fractal dimensions.
To use an analogy from the clusters in Section \ref{clusters}, larger 
structures begin overlapping before smaller structures, with the result
that, {\em for fractal dimensions $F_{\rm dim}<3$, the MSDC is more sensitive to 
the depth of a SFR on large scales than on small scales}.  In particular,
for a fractal dimension of $F_{\rm dim}=1.4$, the MSDC on scales less than $\approx
1/25$ of the survey region's dimensions varies by less than a factor 
of 2 even if the depth is increased by a factor of 100.

\section{The break between the binary and large-scale regimes}
\label{posbreak}

Larson \shortcite{Larson95} demonstrated that there is a break in the 
slope of the MSDC for the Taurus-Auriga SFR, and Simon \shortcite{Simon97} 
found similar breaks in the MSDCs of the Ophiuchus and Orion Trapezium 
regions.  Larson identified the
break as being at the length scale where the MSDC goes from being dominated
by binaries to being dominated by large-scale structure (e.g. clustering
or self-similar structure).  
The fact that a break occurs indicates that a single 
scale-free process is not responsible for the formation of both binaries
and large scale clustering.  Larson also noted that the break occurred 
at a length scale roughly equal to the typical Jeans length in the 
Taurus-Auriga SFR and speculated that this was evidence that,
for length scales less than the Jeans length,
binaries are formed via fragmentation of a single collapsing cloud core
while, on larger scales, stars are distributed self-similarly.  However,
Simon \shortcite{Simon97} found that the breaks between the binary and 
large scale regimes occur at smaller separations in star-forming 
regions with higher stellar densities.  Furthermore, in the SFRs considered
by Nakajima et al. \shortcite{NTHN98}, the location of break varied 
from $\approx 1000$ -- 30000\,AU\@.

In this section we consider the origin of the break between the 
binary and large-scale regimes in the MSDC\@.  First, we investigate what 
determines the break in SFRs with a uniform stellar distribution.  
We then consider the behaviour of the transition separation in regions
where the stars are clustered or distributed self-similarly.  Finally,
we predict how the location of the break varies as a SFR evolves with time
due to expansion and/or the erasing of initial self-similar structure.

\subsection{Uniform stellar distributions}
\label{posbreakSFR}

\begin{figure}
\vspace{0.2truecm}
\vspace{8.5truecm}
\caption{\label{posbreak1} The mean surface density of companions $\Sigma_{\rm com}(r)$ for three distributions stellar systems (shown above).  In all cases, each stellar system is a binary system with separation in the range $0.1$ to $10^4$ AU, and the mean three-dimensional distance between the stars is the same ($d_3 = 0.28$ pc = $5.7\times 10^4$ AU, dotted vertical line), but the depth of the volume varies: a) 0.28 pc (top left, and solid line), b) 2.8 pc (top middle, and dotted line), and c) 28 pc (top right, and dashed line).  The break between the binary and large-scale regimes moves to smaller separations simply because the projected surface density of stars increases.}
\end{figure}

Consider $N$ stellar systems with a uniform spatial distribution in a volume
with dimensions on the sky of $X$, $Y$, and depth $Z$.  Those systems that 
are binaries
have separations distributed as in equation \ref{eqsepdist}.  On large
scales the MSDC is simply the mean surface density of stars
\be
\label{equniformMSDC}
\Sigma_{\rm mean} = \frac{N(1+B_{\rm freq})}{XY} = \frac{4}{\pi {d_2}^2},
\ee
where $d_2$ is the mean (two-dimensional) separation of stars on the sky.  
On small scales most companions are members of the binary systems, so that 
the MSDC goes as equation \ref{eqbinaryMSDC}.  The transition between the two
occurs at the separation where the two sections of the MSDC are equal 
(see Figure \ref{posbreak1}).  Equating equations \ref{eqbinaryMSDC} 
and \ref{equniformMSDC} we find that the break occurs at the separation
\be
\label{eqposbreak1}
R_{\rm b} = d_2~\sqrt{\frac{B_{\rm freq}}{8~{\rm ln}[R_{\rm max}/R_{\rm min}]}}.
\ee
The mean separation between the stars on the sky, $d_2$,
depends both on the mean volume density of stars
\be
\label{volumedensity}
\rho = \frac{N(1+B_{\rm freq})}{XYZ} = \frac{6}{\pi {d_3}^3},
\ee
where $d_3$ is the mean three-dimensional distance between stars,
and on the depth $Z$ since
\be
\label{eqsigmarho}
\Sigma_{\rm mean} = \rho Z.
\ee
Thus, using equations \ref{equniformMSDC} to \ref{eqsigmarho},
\be
\label{eqposbreak}
R_{\rm b} = d_3~\sqrt{\frac{d_3}{Z}}~\sqrt{\frac{B_{\rm freq}}{12~{\rm ln}[R_{\rm max}/R_{\rm min}]}}.
\ee
Hence, the separation at which the break between the binary and 
large-scale regimes occurs
depends on the mean volume density of stars, the depth of the SFR 
($Z/d_3$), and,
to some extent, on the parameters of the binaries themselves (the
binary frequency $B_{\rm freq}$ over the separation range $R_{\rm min}$ 
to $R_{\rm max}$).

\begin{figure}
\vspace{0.2truecm}
\centerline{\hspace{0.0truecm}\psfig{figure=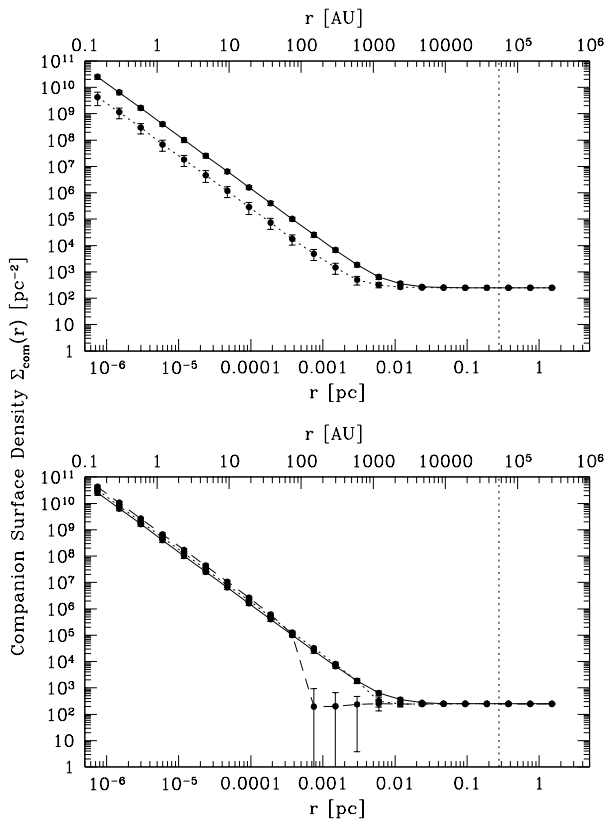,width=11.0truecm,height=11.0truecm,rwidth=9.0truecm,rheight=11.0truecm}}
\caption{\label{posbreak2} The mean surface density of companions $\Sigma_{\rm com}(r)$ for stellar systems distributed as in Figure \ref{posbreak1}b, but with different assumptions about the binaries.  In the standard case (solid lines) each stellar system is a binary system with separation in the range $0.1$ to $10^4$ AU.  In the upper graph a), we also show the effect if, rather than every system being a binary ($B_{\rm freq}=1.0$), only 10 percent of systems are binaries ($B_{\rm freq}=0.1$) with the total number of stars being the same (dotted line).  In the lower graph b), the binary frequency is always $B_{\rm freq}=1.0$, but the upper cut-off in the distribution of separations is varied: 10000 AU (solid line), 1000 AU (dotted line), and 100 AU (dashed line).  The dotted vertical lines give the mean three-dimensional distance between stars, $d_3$.  The break between the binary and large-scale regimes moves to smaller separations if the binary frequency is lower, and there is an abrupt drop in the MSDC if the binary separations are truncated at separations less than value of $R_{\rm b}$ predicted by equation \ref{eqposbreak}.}
\end{figure}

The dependencies of $R_{\rm b}$ are demonstrated in Figures \ref{posbreak1}
and \ref{posbreak2}.  In Figure \ref{posbreak1}a, 50 binary stellar systems 
(100 stars) with separations ranging from 0.1 to $10^4$\,AU
are placed within a volume with dimension on the sky of 2 pc by 2 pc, and
depth $Z=d_3=0.28$ pc (obtained by setting $Z=d_3$ in equation 
\ref{volumedensity}) so that the volume is essentially two-dimensional.
In Figures \ref{posbreak1}b--c, the volume density is kept constant, but
the depth is increased by a factor of 10 each time 
($Z=2.8$ pc and $Z=28$ pc respectively).  This is also equivalent to 
keeping $Z$ fixed, but increasing the volume density by a factor of 10 
each time (except this would mean that wide binaries begin to overlap
with neighbouring systems).  The resulting MSDC for each of
Figures \ref{posbreak1}a--c is
given below the stellar distributions.  
As predicted by equation \ref{eqposbreak},
the position of the break moves to smaller separations with increasing 
surface density, even though the parameters of the binaries remain unchanged.
This explains the finding by Simon \shortcite{Simon97} that the position
of the break varies between star-forming regions and seems to depend on 
the stellar density of the region.

Figure \ref{posbreak2} shows the effects of changing the parameters of the
binaries.  Decreasing the binary frequency, but keeping the total number
of stars the same moves the break to smaller separations.  Keeping the
binary frequency constant, but decreasing the range of separations moves the
break to larger separations, so long as the maximum binary separation is 
greater than $R_{\rm b}$.  These effects are minor, however, compared
to changes in the large-scale surface density.

While Larson \shortcite{Larson95} identified the break in the MSDC of the
Taurus-Auriga SFR with the typical Jeans length, $R_{\rm Jeans}$, in the
molecular cloud,
equation \ref{eqposbreak} shows that the separation at 
which the break occurs is {\em not necessarily} equal to the Jeans length.
The basis of Larson's argument that the break identifies the Jeans length
is that binaries form from the collapse and fragmentation of 
Jeans-critical cloud cores, while
large-scale structure depends on the structure in molecular clouds.  From 
this, one may expect that the maximum binary separation is approximately equal
to the size of the cores and, thus, is less than or approximately equal to
the mean distance between stars $d_3$.  However, we found above that the
break occurs when a {\em projected} companion has an equal probability of 
being a binary or non-binary companion.  
From equation \ref{eqposbreak} and Figures \ref{posbreak1} and
\ref{posbreak2}, for reasonable binary parameters
and $Z \simgreat d_3$, the separation at which the break
occurs is {\em always less} than $d_3$ by at least an order of magnitude
(dotted vertical lines), {\em regardless} of the Jeans length.
Also, binaries may exist with separations 
$\gg R_{\rm b}$ (especially if $Z > d_3$); they are just hidden because 
the mean surface density of stars in the survey region is greater 
than the MSDC of binary companions at large separations 
(i.e. the confusion limit has been reached).  Thus, {\em in general,
for a uniform distribution of stellar systems, the break in the MSDC 
function does not give the Jeans length in a
star-forming region}.  For $R_{\rm b}$ to be similar to the 
Jeans length, either the star-forming region must be very shallow 
(i.e. $Z \simless d_3$) in which case $R_{\rm b} \approx d_3 
\approx R_{\rm Jeans}$, or $R_{\rm Jeans} \ll d_3$ in which case
$R_{\rm b}$ might approximate the Jeans length because $R_{\rm b} \ll d_3$
also.

\subsection{Non-uniform stellar distributions}
\label{nonuniform}

To obtain equation \ref{eqposbreak}, we assumed the stellar systems had
a uniform spatial distribution.  For non-uniform stellar distributions, 
the separation at which the transition from the binary to the
large-scale regimes occurs has the same dependence on the parameters 
of the binaries.
However, rather than the break occurring when the MSDC of the binaries 
(equation \ref{eqbinaryMSDC}) is equal to the mean surface density of stars
(equation \ref{equniformMSDC}), the break occurs at the separation where
equation \ref{eqbinaryMSDC} is equal to the MSDC of the distribution
of stellar systems, which may have its own structure (e.g. Sections 
\ref{globaldensitygradient}, \ref{globalcondensed}, \ref{clusters}, 
\ref{selfsimilar} and \ref{taurus}).
If the volume-filling factor of stellar systems
is high in the survey volume, the transition separation still depends on
the volume density of stellar systems, and on the depth of the survey volume.
However, if the volume filling factor is low the depth of the stellar
distribution may be less important.  To illustrate this point, we consider
the MSDCs both of randomly-distributed clusters of binary stellar systems, 
and of fractal distributions of binary stellar systems.

Consider a stellar distribution comprised of randomly-distributed
clusters of binary stellar systems.  The large-scale MSDC is similar to 
those given in Figure \ref{cluster1}, 
where distributions b) and c) are the same
as a) but the depth of the volume is increased by factors of 10 and 100,
respectively.  Assuming that the MSDC of the binaries equals the MSDC of
the systems at a separation less than the radii of the clusters then,
even though
the depth of the volume is increased by a factor of 10--100, the transition
separation is almost unchanged because the system MSDC on scales less
than the cluster radius is given by the surface density of stars in the
clusters which is unchanged until clusters begin to overlap with each other.

\begin{figure}
\vspace{0.2truecm}
\centerline{\hspace{0.2truecm}\psfig{figure=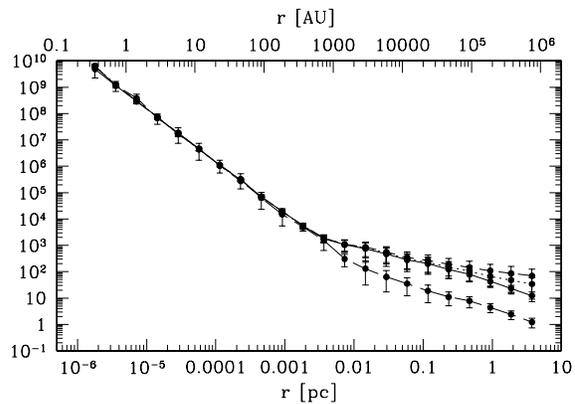,width=11.0truecm,height=11.0truecm,rwidth=9.0truecm,rheight=5.5truecm}}
\caption{\label{fractal2} The mean surface density of companions $\Sigma_{\rm com}(r)$ for stellar systems distributed self-similarly with fractal dimension $F_{\rm dim}=1.4$ (similar to Figure 9a).  Each stellar system is a binary system with separation in the range $0.1$ to $10^4$ AU.  Four different cases are illustrated.  Case a) (solid) has 500 systems (1000 stars) distributed within a region of size 10x10x10 pc (i.e. depth = 1).   Cases b) and c) are identical, except that the depths are increased to 100 (dotted) and 1000 (short-dashed) pc,
respectively.  Case d) is identical to the first, except that only 1/10 of the {\em systems} are included to mimic incompleteness (long-dashed, lower curve).  Note that with $F_{\rm dim}=1.4$, the depth of the survey region has no effect on the location of the break between the binary and the large-scale regimes, while incomplete surveying of stellar systems results in a shift of the break to larger scales.}
\end{figure}

\begin{figure}
\vspace{0.2truecm}
\vspace{8.5truecm}
\caption{\label{fractal3} The mean surface density of companions $\Sigma_{\rm com}(r)$ for four different distributions of 500 stellar systems (three shown above).  Each stellar system is a binary system with separation in the range $0.1$ to $10^4$ AU.  In each case, the stellar systems were initially distributed self-similarly with $F_{\rm dim}=1.4$ and the depth of the region is unity.  However, these distributions were evolved by moving each {\em system} in a random direction by a) 0.01 pc (solid), b) 0.1 pc (dotted line; left), c) 1 pc (short-dashed; centre) and d) 10 pc (long-dashed; right).  This mimics the effect of the distribution evolving with time due to a random system velocity dispersion of $1$ km/s over $10^4$, $10^5$, $10^6$ and $10^7$ years, respectively.  Note that the binary and very-large-scale regimes are unchanged, but at intermediate separations $\Sigma_{\rm com}(r)$ is flat.}
\end{figure}

Figure \ref{fractal2} gives the MSDC for fractal distributions of binary
stellar systems.  In each case, the fractal dimension is $F_{\rm dim}=1.4$.  In cases
b) and c) the depth of the region is 10 or 100 times that of case a), 
respectively.  As expected, with such a low fractal dimension, 
the transition separation is almost independent
of the depth (c.f. Figure \ref{fractal1}).  
A variation in the transition separation can be obtained if the 
number of systems that are included in the calculation of the MSDC 
is decreased (Figure \ref{fractal2}d, which mimics an incomplete survey) or if
the dimensions of the survey region are changed but the number of systems 
is unchanged (i.e. a SFR with a different volume density of stars).  
The difference between Figures \ref{fractal2}a and \ref{fractal2}d
demonstrates the importance of having a complete stellar survey.

Gomez et al. \shortcite{GHKH93} interpreted the stellar systems in 
the Taurus-Auriga SFR as being primarily grouped in small clusters which were
well-separated on the sky (see also Section \ref{taurus}).
Larson \shortcite{Larson95} interpreted the systems as being distributed
with a fractal dimension $F_{\rm dim}=1.4$.  
In either case, the volume-filling 
factor is low.  Therefore, the depth of the SFR is relatively unimportant 
and the transition separation between the binary and 
large-scale regimes is determined either by the surface density of the stars 
within the clusters, or by the projected separation at which binary
companions are as common as fractal companions.  In the case of the
stars being distributed in clusters, 
when the mean separation of stellar systems is estimated for the stars within 
these clusters (Section \ref{taurus}), the clusters are
found to be approximately `two-dimensional', in that the depth of the clusters
is approximately equal to the mean separation between systems 
($Z \approx d_3$).
Therefore, from equation \ref{eqposbreak}, the transition separation will be
within an order of magnitude of $d_3$ and hence similar to the maximum
binary separation and presumably to the size of Jeans-critical cloud cores.  
For the fractal distribution, projection effects are unimportant due to the
low fractal dimension and, therefore, the projected separation at which 
binary companions are as common as fractal companions is approximately 
equal to the three-dimensional separation at which binary and fractal
companions are equally likely.  Thus, as with the clusters interpretation,
the transition separation 
again closely estimates the typical size of Jeans-critical cloud cores.
This explains why Larson \shortcite{Larson95} obtained good agreement
between $R_{\rm b}$ and the expected Jeans length in the Taurus-Auriga SFR.
In regions such as the Orion Trapezium Cluster, however, this is not the
case since the volume-filling factor is high and the SFR is deep 
(Section \ref{otc}).

\subsection{Evolution with time}
\label{evoltime}

\subsubsection{Erasure of substructure}
\label{erasure}

If the stellar distribution in a star-forming region has some 
structure initially, this structure will slowly be lost as the system
evolves due to the stellar velocity dispersion and the evaporation of bound
clusters.  Structure will be lost on the smallest scales first, with larger
scales being affected as the SFR ages.  In the long-term, especially
for SFRs that are initially unbound, the stars will evolve towards a 
homogeneous distribution with no structure (i.e. a flat MSDC).

Assuming that self-gravity is unimportant in maintaining initial structure
(e.g. unbound associations rather than bound clusters), the effects of
time can be mimicked by moving each stellar system (single, binary, or
higher-order multiple system), in a random direction, by the
mean system velocity dispersion multiplied by the time since the stars formed.
In Figure \ref{fractal3}, we show how an initially fractal distribution of
binary stellar systems evolves with time due to such a velocity dispersion.
As predicted, fractal structure on the smallest scales is erased first and the
distribution becomes homogeneous on those scales.  For an initially fractal
distribution, this results in a flat MSDC on scales larger than the
binary regime and smaller than the undisturbed fractal regime, with the
homogeneous regime extending to larger length scales with time.  Eventually,
over a finite area, the stars have mixed so much that any initial structure
is completely lost (after $10^7$ years in Figure \ref{fractal3}).  
Note also that the break between the binary and 
large-scale regimes moves to larger separations with time since the smallest
structures (which have the highest stellar density) are disrupted first.
When all structure has been erased, the transition separation is simply given 
by equation \ref{eqposbreak}.  Finally, the systems will continue to disperse
after all the initial structure has been erased, further lowering the 
MSDC on large-scales, and thus the transition separation will continue to
grow (see also Section \ref{expansion}).

A similar process occurs with unbound clusters (associations)
of stars; the clusters get larger in spatial extent, their
stellar density decreases, and hence the break between the binary and 
large-scale regimes moves to larger separations.  

With bound clusters, structure on the smallest scales will again
be erased rapidly, and the only
remaining structure will be the radial density profile of the overall
cluster which, as seen in Section \ref{globalcondensed}, has a very flat
MSDC (a singular isothermal sphere has an MSDC with a power-law slope of 
$\approx -0.2$, and less centrally-condensed configurations are even flatter).

\subsubsection{Expansion of open clusters}
\label{expansion}

In the above sections, considering the 
derivation of the position of the break between the 
binary and large-scale regimes, we have assumed that binaries with 
separations $\ge R_{\rm b}$ exist.  Although this is likely in a young 
star-forming region, it may not always be the case.  
For example, consider the evolution of an open
cluster.  When the cluster has just formed, the mean distance between stellar
systems is small and wide binaries are prone to being disrupted (e.g.
Kroupa 1995a,b,c).  Later in the cluster's evolution
it expands due to gas loss.  This lowers the mean density of stars and 
hence the break between the binary and large-scale regimes in the MSDC
moves to larger separations.  However, since wide binaries
have been destroyed, the predicted position of the break may exceed the
separations of the widest binaries.  The result is that
the MSDC shows the characteristic slope of $\approx -2$ for small separations
where the binaries exist, but that at the separation beyond which binaries 
have been destroyed, the MSDC drops abruptly to the mean surface density
of stellar systems within the survey region (e.g. Figure \ref{posbreak2}b).
Such absence of wide binaries may be visible in open clusters such
as the Pleiades and Hyades.  If so, this would give an indirect record 
of how dense the star clusters were when the stars were initially formed: 
knowing the widest binaries remaining and assuming a timescale for the
expansion of the cluster from its initial density, the stellar density 
required to destroy wider binaries on a timescale less than the expansion 
time of the cluster can be determined (see Section \ref{orionbinary}; 
Binney \& Tremaine 1987).

Finally, note that for separations larger than that of the widest binary the
companions to stars from which the MSDC is calculated are given by chance
projections or unbound neighbours.  In a survey of a small number of stars
it is likely that no such companions exist which results in a gap in the
MSDC rather than a drop from the binary regime to the mean surface density
of stars (hence the large errorbars for separations just greater than the 
maximum binary separation in Figure \ref{posbreak2}b).

\section{Application to star-forming regions}
\label{application}

In the previous sections, we have examined the mean surface density of 
companions of global density profiles, binaries, clusters, fractals and have 
determined the dependencies and evolution of the position of the break 
between the binary
and large-scale regimes.  We now apply these results to the Taurus-Auriga,
Ophiuchus, and Orion Trapezium star-forming regions.  Taurus-Auriga has been 
studied by Larson \shortcite{Larson95}, while Simon \shortcite{Simon97} 
studied all three regions and Nakajima et al. \shortcite{NTHN98} studied
Ophiuchus and the Orion A, B, and OB star-forming regions.
However, in light of the above results, a reanalysis is worthwhile.

\subsection{Taurus-Auriga}
\label{taurus}

\begin{figure}
\vspace{0.2truecm}
\centerline{\hspace{0.0truecm}\psfig{figure=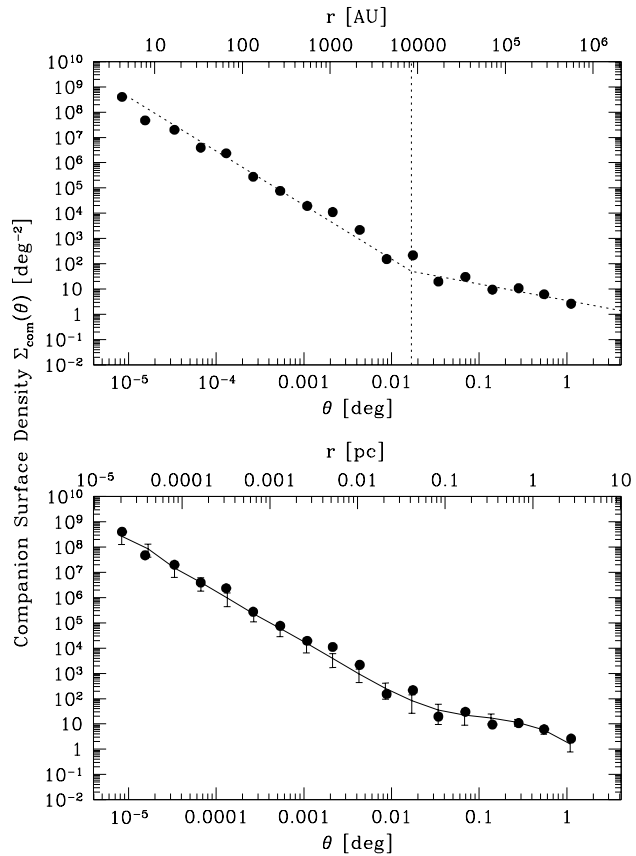,width=11.5truecm,height=11.5truecm,rwidth=9.0truecm,rheight=11.5truecm}}
\caption{\label{taurus1} The mean surface density of companions $\Sigma_{\rm com}(r)$ for the Taurus-Auriga SFR.  In the upper graph a), the MSDC derived by Simon (1997) (points) is overlaid on the power-law fit to the MSDC obtained by Larson (1995).  In the lower graph b), the MSDC of binary stellar systems distributed in randomly-positioned clusters (see text) is overlaid on the MSDC derived by Simon (1997).  Both the broken power-law and cluster-MSDC provide good fits to the data, indicating that the interpretation of the sloping MSDC on large scales being due to fractal structure is not the only interpretation.}
\end{figure}

Both Larson \shortcite{Larson95} and Simon \shortcite{Simon97} determined
the MSDC for the Taurus-Auriga SFR\@.  Larson used the sample of 121 sources
from Gomez et al. \shortcite{GHKH93} for the large-scale distribution of 
stellar systems ($\theta > 5''$) combined with data from
Simon \shortcite{Simon92}, Ghez et al.
\shortcite{GheNeuMat93}, and 
Leinert et al. \shortcite{LZWCRJHL93} for small separations 
($0.1 \simless \theta < 10''$).  
Simon \shortcite{Simon97}, on the other hand, 
used a sub-sample of just 49 systems that had been well-surveyed for 
companions in the separation range 0.005 to $10''$ resulting in a total of
80 stars (Simon et al. 1995).

This difference in samples is important, since it is relatively meaningless
to study the MSDC of an incomplete and/or non-uniformly sampled fraction of
the total stellar population of a region.  
The Gomez et al. \shortcite{GHKH93} sample was relatively complete to
systems with minimum separations $\approx 5''$,
and as a consequence, they stated in their paper: 
``... most of the optically visible objects
with $V\simless 15.5$ close to the Taurus-Auriga molecular cloud have already 
been found.  ...heavily extincted infrared sources ... appear to constitute
a modest fraction of the total pre-main-sequence population.  Therefore it
seems worthwhile to begin to investigate the spatial distribution of young
stars in Taurus.''

Thus, the cumulative sample used by Larson \shortcite{Larson95} gives a better
determination of the MSDC on the large scales than the incomplete
sub-sample used by Simon \shortcite{Simon97}, 
while on the small scales, the two
samples are fairly similar, since most of the systems used
by Simon were previously known and used by Larson. 

Both Larson and Simon found that the large-scale regime can be well fit
by a power-law slope of $\approx -0.6$ (see Figure \ref{taurus1}a), which
both took to imply a self-similar or fractal distribution.
However, we have shown in Section 2 that 
a given MSDC slope does not correspond to a unique
density distribution, and thus it is interesting to ask
how robust the result of Larson and Simon is. For example,
rather than requiring a self-similar structure to fit the
data, could the stars simply be in randomly-distributed
clusters?

In Figure \ref{taurus1}a, we show the MSDC for Taurus-Auriga derived by
Simon (points) with the split power-law fit derived by
Larson (dotted line). In Figure \ref{taurus1}b, the same MSDC data
from Simon are again plotted as points, but the solid line
fit is now from a model stellar distribution.  We took 
40 stellar systems, each composed of a binary
distributed in the separation range 1 -- $2 \times 10^4$ AU 
according to equation \ref{eqsepdist}.  Note that since Simon defined two
stars to be a binary if their separation was $\theta < 10'' \approx 1400$ AU,
our choice of 40 binary systems with a larger range of separations gives
roughly the same numbers of stars (80) and systems (49) that are contained 
in Simon's sample.

The stellar systems are distributed
over the same area as Simon's sample of stars,
with 30\% randomly distributed, and the
remaining 70\% in clusters of radius $R_{\rm clus} = 2$ pc,
each with 8 systems (16 stars) centrally condensed with
$\rho \propto r^{-2}$.  We assume a distance of 140 pc to 
Taurus-Auriga \cite{WBKJR98}.  Such a distribution is consistent
with that determined for the Taurus-Auriga SFR by Gomez
et al. \shortcite{GHKH93} since, although they find clusters consisting 
of 9--18 stars (with separations greater than $\approx 5''$) 
with radii $0.5-1.1$ pc, the radii depend on the arbitrary
stellar surface density at which the cluster is defined to stop.  We point
out that the mean distance between systems in each cluster is $\sim 1$ 
pc, and therefore each cluster is essentially two-dimensional 
(i.e.~$R_{\rm clus} \approx Z \approx d_3$; see 
Sections \ref{posbreakSFR} and \ref{nonuniform}).

This model MSDC provides an excellent fit to the data points
of Simon (Figure \ref{taurus1}b) 
and, moreover, it is not unique, with other randomly-distributed 
cluster models possible. Thus, while the large-scale
structure in Taurus-Auriga is consistent with a fractal structure
of dimension 1.4 (Larson 1995; Simon 1997), it is equally
consistent with random clustering.  To differentiate between
the two would require extending the MSDC to larger scales:
however, this is not possible, since the data used by Larson
already covers the full star-forming region.  

Another way to differentiate
between the two might be to use higher-order correlation functions
than the MSDC (which is equivalent to the two-point correlation function)
and/or the nearest-neighbour distribution.  The latter has been used by
Nakajima et al. \shortcite{NTHN98} to argue that a MSDC with a power-law
slope that is negatively large might indicate a spread of stellar ages
rather than stars being formed with a self-similar distribution.  They
model the Orion B SFR by a mixture of randomly-distributed stars and clusters
of different stellar surface densities (the larger of which are assumed to
be older
than the smaller ones) and obtain a good fit to the MSDC
and a better fit to the nearest-neighbour distribution than that implied by 
a purely fractal stellar distribution.  This is similar to the above
demonstration that the Taurus-Auriga MSDC can be modelled using 
clusters rather than fractal structure (although we use several identical
clusters rather than a mixture).  However, we point out that the
nearest-neighbour distributions found by Nakajima et al. \shortcite{NTHN98}
{\em do not exclude} the possibility that stars are formed with a 
self-similar distribution, since the
nearest-neighbour and MSDC distributions can be fit by the combination of
fractal and homogeneous (presumably older) populations as well as
the cluster and homogeneous populations of Nakajima et al.


\subsection{Ophiuchus}

Simon \shortcite{Simon97} studied the spatial distribution of 35 
stellar systems (51 stars) in the Ophiuchus SFR.
He found that on small scales ($\simless 5000$ AU) 
the MSDC had a power-law slope of $-1.9 \pm 0.1$
while on large-scales the MSDC could be described by a power-law
slope of $-0.5 \pm 0.2$ (similar to the Taurus-Auriga SFR).  Nakajima et al.
\shortcite{NTHN98} studied two samples of stars from the Ophiuchus SFR.
The first sample consisted of 86 H$\alpha$
emission line stars (Wilking, Schwartz \& Blackwell~1987), 
10 of which are known 
binaries \cite{ReiZin93}.  The second sample
consisted of 78 embedded stars detected in the infrared by 
Wilking, Lada, \& Young \shortcite{WilLadYou89}
(18 of which are in common with 
the first sample).  They found the MSDC of the H$\alpha$ sample had a 
power-law slope of $-2.5 \pm 0.3$ on small scales ($\simless 5000$ AU)
and a slope of $-0.36 \pm
0.06$ on large scales.  The embedded sample had a slope of $-0.28$ on 
large scales and combining the samples gave a slope of $-0.26$ on large
scales.

The slopes on small scales are consistent with binaries having an 
approximately uniform distribution
of separations with the logarithm of separation.  However, as 
described in Sections \ref{nonuniform} and \ref{taurus}, when 
studying the large-scale distribution of stars
in a SFR, it is essential to have the positions of the majority of the young
stars in the survey region.  Unfortunately, as demonstrated by the three
different samples used by Simon \shortcite{Simon97} and Nakajima et al.
\shortcite{NTHN98}, there is no complete sample of young 
pre-main-sequence stars available for Ophiuchus at this time 
(unlike for Taurus-Auriga).
Another survey is that of Strom, Kepner, \& Strom 
\shortcite{StrKepStr95} who find 119 pre-main-sequence stars, 
in three rich aggregates of young stars in the Ophiuchus cloud along with
a more distributed population.  Therefore, we prefer to wait until a 
more comprehensive list of young pre-main-sequence stars is available.

\begin{figure}
\vspace{-0.2truecm}
\centerline{\hspace{0.2truecm}\psfig{figure=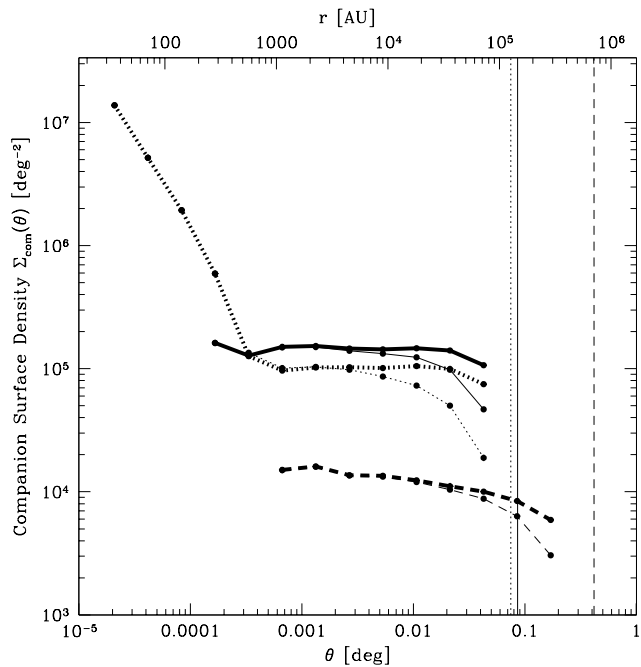,width=9.0truecm,height=9.0truecm,rwidth=9.0truecm,rheight=9.0truecm}}
\caption{\label{orion1} The mean surface density of companions $\Sigma_{\rm com}(\theta)$ for the stellar distributions from the surveys of the Orion Trapezium Cluster by Jones \& Walker (1988) (dashed lines), Prosser et al. (1994) (dotted lines) and McCaughrean et al. (1996) (solid lines).  In each case, the MSDC is calculated using Method 1 (thin lines), ignoring the effects of boundaries, and using Method 5 (thick lines) which attempts to correct for the boundaries of the survey regions.  The vertical lines give the smallest dimension of the survey regions.}
\end{figure}

There are several comments to be made about the current analyses, however.
First, the power-law slopes on large scales found by Nakajima et al. 
\shortcite{NTHN98} are quite flat and may be even flatter since 
they do not take into account boundary effects when
calculating the MSDC (see Section \ref{otc}).  Second, for the 
embedded stars the power-law slope is calculated over separations from 
$\approx 3000 - 65000$ AU.  A flat MSDC on these scales is not unexpected
since with a velocity dispersion of 1~km~s$^{-1}$, any initial structure
would be erased on these scales in $\approx 3 \times 10^5$ years.
Finally, the effects of the evolution of the MSDC with age may be 
apparent in the different samples used by Nakajima et al. \shortcite{NTHN98}.
The H$\alpha$ sample is presumably older than the embedded sample,
is more dispersed, and has a lower mean surface density.  This
clearly results in a nearest-neighbour distribution which is non-Poisson
when the two samples are combined, regardless of whether or not there is
sub-structure in the stellar distributions.  To determine whether or
not there is structure (e.g. sub-clustering or self-similar structure) 
in the distribution of forming stars, it is important to consider the youngest
stars and the largest possible area.  The youngest stars
should be selected to minimise the erasure of structure (Section \ref{erasure})
and to avoid washing out any structure with a more uniform, older population
(e.g. Figure \ref{cluster3}).  A large area allows the large-scale 
MSDC to be studied over the greatest possible range of separations.  
Note also that infrared surveys
are preferred over optical, since they give greater completeness 
and minimise the problem of finding structure that is due to obscuration
of stars by molecular gas rather than structure in the distribution 
of the stellar objects themselves.

\subsection{Orion Trapezium Cluster}
\label{otc}

Finally, Simon \shortcite{Simon97} studied the spatial distribution of the
Orion Trapezium Cluster, using the sample of 319 stars listed by 
Prosser et al. \shortcite{PSHSJWM94} from their optical wavelength Hubble
Space Telescope study which probed for binaries down to $\approx 0.05''$ or
$\approx 25$ AU separation.  As with the Taurus-Auriga
and Ophiuchus SFRs, Simon \shortcite{Simon97} 
found the MSDC on small-scales ($\simless 1''$) is 
consistent with binaries having an approximately uniform distribution
of separations with the logarithm of separation (with a power-law
slope of $-2.1 \pm 0.6$).  On large-scales (2.5 to $40''$), Simon 
found the MSDC could be described by a power-law slope of $-0.2 \pm 0.2$.
Nakajima et al. \shortcite{NTHN98} studied the whole Orion A region and
found a similar slope ($-0.23 \pm 0.02$) to Simon.  However, the Orion A
region has a much larger spatial extent than the Orion Trapezium 
Cluster itself and therefore is not directly comparable to Simon's results
or those presented here.

Figure \ref{orion1} gives the MSDC for the data of Prosser et al. 
over the full range of separations, calculated using Method 1 (thin, dotted
line) and Method 5 (thick, dotted line).  
The MSDC for separations $\simgreat 10''$ is entirely dependent
on the corrections applied because of the boundaries of survey area.  Fitting
a power-law slope to the MSDC given by Method 1 over the same large-scale
range as Simon ($\approx 3 - 50''$) gives a slope of $-0.17 \pm 0.03$ in
agreement with Simon (we obtain a smaller standard error since we do
not attempt to give error estimates for each MSDC value).
However, using Method 5, the slope is $0.01 \pm 0.01$
over both the range used by Simon and over the larger range of $1.6-100''$.  
Thus, Simon's conclusion that the power-law slope in the large-scale 
regime is the same in the Trapezium Cluster as in Taurus-Auriga appears 
to be based entirely on boundary effects (although we note that a
slope of zero is within $1\sigma$ of Simon's result due to his 
large error estimate).  In a similar manner, since Nakajima et al. 
\shortcite{NTHN98} use the same method as Simon to calculate the MSDC,
their large-scale power-law slopes are also likely to overestimated in
magnitude.

\begin{figure}
\vspace{-0.2truecm}
\vspace{7.5truecm}
\caption{\label{orionfield} The stellar distributions from the surveys of the Orion Trapezium cluster by Jones \& Walker (1988), Prosser et al. (1994) and McCaughrean et al. (1996). }
\end{figure}

To better analyse the Trapezium Cluster, at least
on the larger scales, we must turn to other data sets for
several reasons.  First, the Prosser et al. \shortcite{PSHSJWM94}
data cover a 
relatively limited part of the central cluster, albeit omitting 
the highest-density region including the Trapezium stars themselves. 
Second, the survey was at optical wavelengths, making it possible
that younger, more embedded cluster members were missed. Finally, 
the survey boundary is very irregular making it difficult to 
correct the MSDC for boundary effects and raising the possibility
that, if structure is found on large scales, it may simply be
due to the non-uniform sampling of the cluster.

Figure \ref{orionfield} shows the coverage of the Prosser et al. survey
and two alternate surveys. The first
is that of McCaughrean et al. \shortcite{MRZS96} who imaged the central
$5'\times 5'$ of the cluster at 2$\mu$m. The completeness
limit was $K = 17^{\rm m}$, and the steep turnover in
the luminosity function at $K \sim 12^{\rm m}$ ensures that
the sample of 744 sources is very complete, at least to systems
more widely separated than the 0.7$''$ resolution
limit. The second sample is that of Jones \& Walker \shortcite{JonWal88} who
carried out a ground-based optical photographic survey out
to roughly $15'$ from the centre of the Trapezium,
(i.e. a much larger area than the other two samples). The
sample contains 858 stars with $\geq$90\% probability of
being proper-motion members of the Orion complex. However,
overall it is a less complete sample, with shallower
detection and coarser resolution limits respectively.

The MSDC of all three surveys are given in Figure \ref{orion1}.  The 
agreement between the large-scale MSDCs produced with the data of 
Prosser et al. 
\shortcite{PSHSJWM94} and McCaughrean et al. \shortcite{MRZS96} is good
because the two surveys contain a lot of overlap.  The McCaughrean MSDC is
higher overall primarily due to the inclusion of the high-stellar-density 
region centred on the Trapezium, while the Prosser MSDC reaches to smaller 
separations due to the superior resolution.  Finally, notice that the
boundary correction
has less effect on the McCaughrean MSDC than on the Prosser MSDC because
the former has more uniform coverage than the latter.  
Fitting a power-law slope
to the McCaughrean MSDC over the range $1.6-100''$ gives a slope of 
$-0.11 \pm 0.02$ (Method 1) or $-0.02 \pm 0.01$ correcting for boundary effects
(Method 5).
The MSDC of the Jones \& Walker \shortcite{JonWal88} data is lower 
overall primarily due to the lack of completeness (and thus the lower stellar
surface density), but also because the 
data extends out much farther where the stellar density (even correcting
for completeness) is much lower and this lowers the {\em mean} 
surface density of
companions.  Unlike the Prosser and McCaughrean MSDCs (which are essentially
flat outside the binary regime using Method 5), the Jones \& Walker MSDC 
is fit by a small power-law slope of $-0.17 \pm 0.03$ (Method 1)
or $-0.16 \pm 0.01$ (Method 5) over the range $1.6-400''$.

In previous sections we have seen that a sloping MSDC can result from
many different effects (e.g. boundaries, global density profiles, clusters,
and fractal stellar distributions).  Thus, rather than interpreting the 
Trapezium Cluster 
large-scale MSDCs simply by fitting power-laws, we choose to  
compare the MSDCs with those of model stellar 
distributions.  We assume a distance of 480 pc to the Trapezium
Cluster \cite{GRMD81}.  
Due to the superior datasets in the large-scale regimes,
we only consider the data of McCaughrean et al. \shortcite{MRZS96} 
and of Jones \& Walker \shortcite{JonWal88}.  Also,
from this point on, we use Method 5 to calculate the MSDC,
although since we are modelling the MSDC, rather than deriving slopes, this
choice is unimportant.

\begin{figure}
\vspace{0.2truecm}
\centerline{\hspace{0.0truecm}\psfig{figure=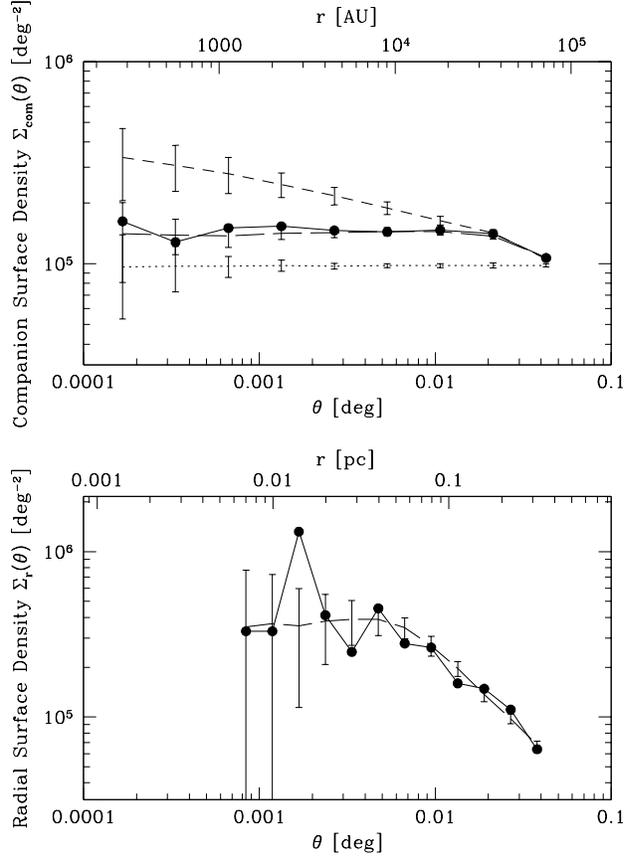,width=11.5truecm,height=11.5truecm,rwidth=9.0truecm,rheight=11.5truecm}}
\caption{\label{mccaughrean} The upper graph a), gives the mean surface density of companions $\Sigma_{\rm com}(\theta)$ for the McCaughrean et al. (1996) survey (points and solid line), along with the MSDC of three models: a uniform distribution of stars (dotted line); a singular isothermal sphere (short-dashed line); and a stellar distribution given by equation \ref{trapfit} (long-dashed line).  The lower graph b), gives the radial surface density $\Sigma_{\rm r}(\theta)$ of the McCaughrean et al. survey along with that of the stellar distribution given by equation \ref{trapfit}.}
\end{figure}

\begin{figure}
\vspace{0.2truecm}
\centerline{\hspace{0.0truecm}\psfig{figure=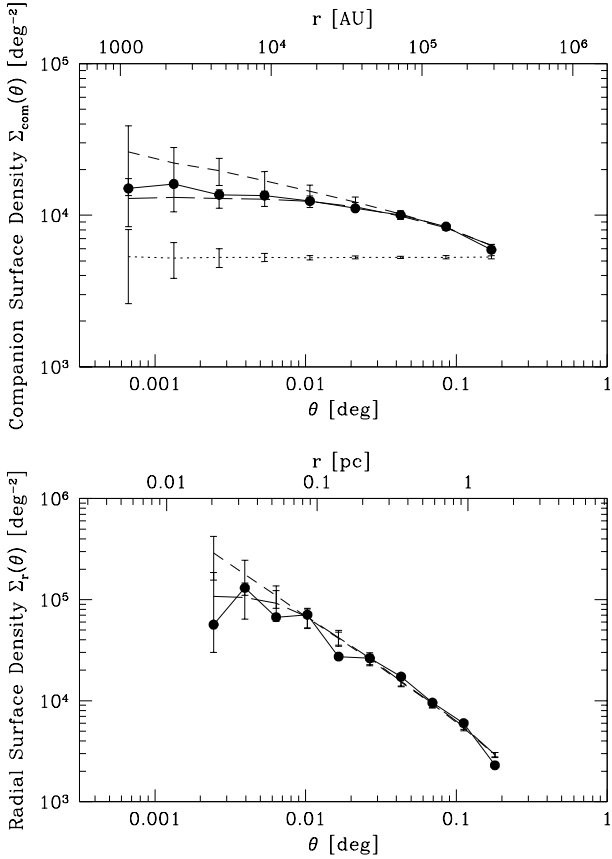,width=11.5truecm,height=11.5truecm,rwidth=9.0truecm,rheight=11.5truecm}}
\caption{\label{joneswalker} The upper graph a), gives the mean surface density of companions $\Sigma_{\rm com}(\theta)$ for the Jones \& Walker (1988) survey (points and solid line), along with the MSDC of three models: a uniform distribution of stars (dotted line); a singular isothermal sphere (short-dashed line); and a stellar distribution given by equation \ref{trapfit} (long-dashed line).  The lower graph b), gives the radial surface density $\Sigma_{\rm r}(\theta)$ of the Jones \& Walker survey along with those of the singular isothermal sphere and the stellar distribution given by equation \ref{trapfit}.}
\end{figure}

Figure \ref{mccaughrean}a gives the MSDC of the McCaughrean et al. data, 
along with the MSDC for three model stellar distributions.  In the first
model distribution (dotted line), 
744 stars are distributed uniformly over the survey 
region.  This results in an MSDC that is too low to fit the Trapezium 
Cluster data.
In the second, 744 stars are distributed randomly with the volume density
distribution of a singular isothermal sphere ($\rho \propto r^{-2}$).
Assuming the depth of the cluster is much greater than the radius of the 
survey region this is equivalent to the radial surface
density distribution $\Sigma_{\rm r}(r) \propto r^{-1}$.
This time the MSDC is too high (short-dashed line), 
and has a slope that is inconsistent with the Trapezium Cluster data.

In Figure \ref{mccaughrean}b, the radial surface density
profile $\Sigma_{\rm r}(r)$ is given for McCaughrean et al. data.
A good fit is obtained (long-dashed line) if the stars of the 
Trapezium Cluster are 
assumed to be distributed with a core of uniform volume density and
$\rho \propto r^{-2}$ outside the core
\be
\label{trapfit}
\rho(r) = \cases{
\rho_0 & if $r \leq R_{\rm core}$, \cr
\rho_0 \left(\frac{R_{\rm core}}{r}\right)^2 & if $r>R_{\rm core}$ \cr }
\ee
with $R_{\rm core} = 30''$ and assuming that the depth of the 
three-dimensional 
distribution is much greater than the radius of the survey region.  With 744
stars distributed within the survey area of McCaughrean et al. this
gives a stellar density of $\rho_0 = 2.1 \times 10^4$ pc$^{-3}$ 
in the central core which is in good agreement with the value of 
$1.7 \times 10^4$ pc$^{-3}$ derived by Hillenbrand \& Hartmann 
\shortcite{HilHar98} who fit a King model to the cluster, but less than
the estimate of $5 \times 10^4$ pc$^{-3}$ by McCaughrean \& Stauffer 
\shortcite{McCSta94} who fit the cluster with a King model with a smaller
core radius.
The MSDC of this stellar distribution is in excellent agreement with 
that of the Trapezium Cluster data (Figure \ref{mccaughrean}a).  Thus, the 
distribution of stars in the central $5'\times 5'$ of the 
Trapezium Cluster
can be well described by this approximate non-singular isothermal sphere.
In particular, {\em there is no evidence of 
sub-clustering or fractal structure}.  Note also, that stellar densities
given by this model and the lack of structure in the stellar distribution
make the cluster three-dimensional in the sense of Section \ref{posbreakSFR}
(i.e. $Z \gg d_3$).  Thus, the break between the binary and the large-scale
regimes in the Prosser MSDC cannot be associated with the typical Jeans 
length, even if the stellar 
distribution had not evolved since the stars formed 
(see Section \ref{orionsubcluster}).

\begin{figure}
\vspace{0.2truecm}
\vspace{13.5truecm}
\caption{\label{orionclus1a} The mean surface density of companions $\Sigma_{\rm com}(\theta)$ for models of the Orion Trapezium Cluster with stars distributed in sub-clusters.  The models are compared to the MSDC of the McCaughrean et al. (1996) survey (points).  In each model, the stars or sub-cluster centres are distributed according to equation \ref{trapfit} and the sub-clusters each contain 10 stars which are distributed uniformly within a sphere of radius $R_{\rm clus}$.  The models have: 100\% of stars in sub-clusters with $R_{\rm clus}=3000$ AU (top left, dotted line);  10\% of stars in sub-clusters with $R_{\rm clus}=3000$ AU (top right, solid line); 100\% of stars in sub-clusters with $R_{\rm clus}=10^4$ AU (lower left, dashed line); 100\% of stars in sub-clusters with $R_{\rm clus}=3\times10^4$ AU (lower right, long-dashed line). }
\end{figure}

\begin{figure}
\vspace{0.2truecm}
\centerline{\hspace{0.2truecm}\psfig{figure=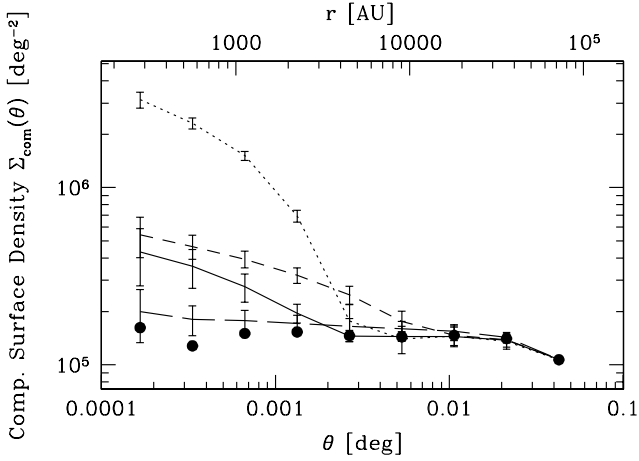,width=13.0truecm,height=13.0truecm,rwidth=9.0truecm,rheight=6.5truecm}}
\caption{\label{orionclus1b} The same as in Figure \ref{orionclus1a} but with centrally-condensed sub-clusters with stars distributed inside the radius $R_{\rm clus}$ according to $\rho \propto r^{-2}$.}
\end{figure}

\begin{figure}
\vspace{0.2truecm}
\centerline{\hspace{0.0truecm}\psfig{figure=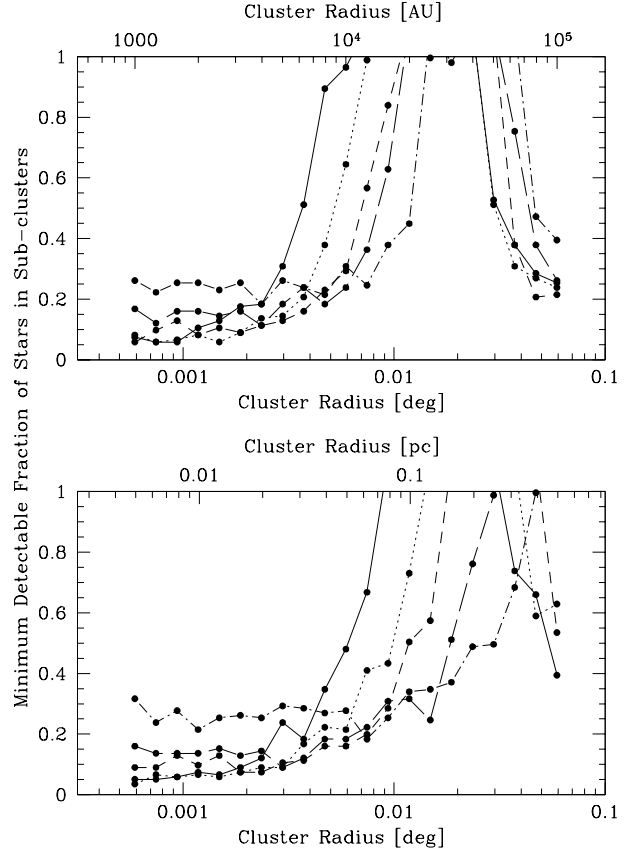,width=11.5truecm,height=11.5truecm,rwidth=9.0truecm,rheight=11.5truecm}}
\caption{\label{orionclus2} The minimum fraction of stars in sub-clusters that are detectable by comparing the MSDC of model stellar distributions to that of the McCaughrean et al. (1996) data.  The minimum fraction that can be detected depends on the radii and number of stars within the sub-clusters and the degree of central condensation of the sub-clusters.  In the upper graph a), the sub-clusters have a uniform stellar density while in the lower graph b), the sub-clusters have a $\rho \propto r^{-2}$ stellar distribution.  The sub-clusters contain 5 (solid lines), 10 (dotted lines), 20 (dashed lines), 40 (long-dashed lines) or 80 (dot-dashed lines) stars.}
\end{figure}

In Figure \ref{joneswalker}a, the MSDC of Jones \& Walker's Trapezium 
Cluster data
is compared to the same three stellar distributions as the McCaughrean et al.
data, except this time 858 stars are distributed over the survey area of
Jones \& Walker.  
Also, the approximation that the depth of the cluster is much
greater than the radius of the survey region is no longer a good one.
Therefore, we assume that the Jones \& Walker survey covered most of the 
Trapezium Cluster and take the three-dimensional radius of the cluster to be
only twice the Jones \& Walker survey radius of $15'$.
As for the McCaughrean MSDC, the uniform-density stellar distribution is 
inconsistent with the Jones \& Walker MSDC\@.  It is more difficult to 
distinguish between the pure singular isothermal sphere and the distribution 
given by equation \ref{trapfit} with the Jones \& Walker MSDC than it was
with the McCaughrean MSDC, but equation \ref{trapfit} does provide a better
fit.  Figure \ref{joneswalker}b gives the radial surface density profile.
Both the singular isothermal sphere and equation \ref{trapfit} 
provide reasonable fits to the radial surface density profile, although again
equation \ref{trapfit} is preferred.  In conclusion, the MSDCs of the 
surveys of McCaughrean et al. and Jones \& Walker give {\em no evidence
for sub-clustering or fractal structure in the Orion 
Trapezium Cluster}.

\subsubsection{Sub-clustering}
\label{orionsubcluster}

Although we have shown that there is no evidence from the MSDC for 
hierarchical structure or sub-clustering of the stars in the Orion Trapezium
Cluster, it is of interest to determine what {\em limits} can be placed 
on the presence of such structure.  By generating model stellar distributions
which contain stellar sub-clusters and comparing their MSDCs to that of
the McCaughrean et al. \shortcite{MRZS96} 
survey, we can set limits on how much sub-clustering can be present.

Figure \ref{orionclus1a} compares the MSDCs of four different stellar 
distributions containing stars in sub-clusters to the MSDC of the 
McCaughrean et al. survey.  
In each case single stars, and/or the centres of sub-clusters,
are distributed randomly according to equation \ref{trapfit}.  A total of
744 stars are allocated, some fraction of which are in sub-clusters.  
Each sub-cluster
consists of 10 stars randomly distributed within the sub-cluster's radius 
$R_{\rm clus}$.  Comparing the models to the McCaughrean MSDC shows that
the ease with which sub-clusters are detected depends on the total fraction 
of the stars that are in such sub-clusters and the radii of the sub-clusters.
For example, the MSDC produced when 100\% of the stars are in
sub-clusters of radii 3000 AU is clearly inconsistent with the 
Trapezium Cluster data (dotted line).
Detecting if 10\% of the stars are in such sub-clusters, however, 
is more difficult (solid line).  
For sub-clusters with radii of $10^4$ AU, any less than
100\% of the stars being in such sub-clusters becomes difficult to detect 
(short-dashed line).
Finally, it is impossible to detect whether or not {\em all} the stars are 
in sub-clusters with radii of $3 \times 10^4$ AU (long-dashed line).
Looking at the examples of each type of clustering in Figure 
\ref{orionclus1a}, we note that the MSDC appears to be slightly less
sensitive than the human eye for detecting sub-clustering, since 
the eye finds it easy to detect clustering in the 
$R_{\rm clus} = 10^4$ AU case while the MSDC detection is only $2-3 \sigma$
(see also Figure \ref{cluster1}).
The advantage of the MSDC over the eye, however, is that it is 
unbiased (the eye is good at seeing patterns where none exist) and
gives a reproducible measure of the sub-structure.

The detection of sub-clustering also depends on the 
central condensation of the
sub-clusters themselves.  In Figure \ref{orionclus1b}, the models are identical
to those in Figure \ref{orionclus1a}, except that the sub-clusters are 
centrally condensed with the 10 stars distributed randomly according to 
$\rho \propto r^{-2}$.  Centrally-condensed sub-clusters are easier to detect
than uniform-density sub-clusters.

In Figure \ref{orionclus2}, we give the minimum fraction of stars that are
required to be in sub-clusters for the resulting MSDC to be inconsistent
with the McCaughrean MSDC\@.  The minimum fraction is a function
of the degree of central-condensation of the sub-clusters,
the number of stars in each sub-cluster (ranging from 5 to 80 stars per 
sub-cluster), and the radii of the sub-clusters (ranging from 1000 to
$10^5$ AU).  For example, in Figure \ref{orionclus2}a, a model with 100\%
of the stars being in sub-clusters of 5 stars (solid line) with
$R_{\rm clus}=2 \times 10^4$ AU gives an MSDC that is indistinguishable 
from the McCaughrean MSDC, whereas if the sub-cluster radii are only 
$R_{\rm clus}=1000$ AU, then a model with as few as 10\% of the stars being in
such clusters can be recognised as being different from the McCaughrean MSDC.
A model MSDC is determined to be inconsistent with the 
McCaughrean MSDC if two or more bins differ from the McCaughrean MSDC values
by more than $2\sigma$.  The easier detection of more centrally-condensed
sub-clusters is apparent (c.f. Figures \ref{orionclus2}a and 
\ref{orionclus2}b).  For uniform-density sub-clusters, strong limits 
($\simless 30$\%) can be placed on the maximum fraction of stars in 
sub-clusters with $R_{\rm clus} \simless 6000$ AU.  For centrally-condensed
sub-clusters, strong limits ($\simless 30$\%) can be placed on the maximum 
fraction of stars in sub-clusters with $R_{\rm clus} \simless 10000$ AU.
The detection also depends on the number of stars in each sub-cluster.
Finally, the possibility that a large fraction of the stars are in 
near-uniform-density sub-clusters with large radii 
($\simgreat 5 \times 10^4$ AU) can be ruled out because the model stellar 
distributions are not as centrally-condensed as the Trapezium Cluster.

Although we cannot rule out all possible sub-clustering in the 
Trapezium Cluster, we emphasise that the McCaughrean MSDC is consistent 
with there being {\em no} sub-clustering.  In fact, this is not too surprising.
The stars in the Trapezium Cluster have a three-dimensional velocity 
dispersion of $\approx 4$ km s$^{-1}$ (Jones \& Walker 1988; Tian et al. 1996).
If the stars formed in unbound associations, or in some type of
hierarchical structure, in $10^5$ years this
velocity dispersion results in the stars typically drifting apart
by $\sim 10^5$ AU.  Hence, any such primordial structure would 
have been destroyed by the current age of the Trapezium Cluster 
($\sim 10^6$ years) \cite{Hillenbrand97}.  Note that this is not the case
with the clustering that is observed in the Taurus-Auriga SFR.  
The smaller velocity dispersion of $1-2$ km s$^{-1}$ \cite{HJSK91}
and large scale of the clustering ($\approx 1$ pc) means that such 
structure takes $\sim 10^6$ years to be destroyed.

On the other hand, if the Trapezium Cluster
stars formed in bound sub-clusters, of $10-100$ stars, there should
still be some evidence.  The maximum lifetime of such a sub-cluster is
determined by its evaporation rate.  For sub-clusters of $10-100$ stars,
the lifetime is $\sim 100 t_{\rm cross}$ \cite{BinTre87}, where $t_{\rm cross}$
is the crossing time.  This gives the lifetime $t$ of a sub-cluster of 
total stellar mass $M_{\rm clus}$ and radius $R_{\rm clus}$ to be
\be
t \sim 3 \times 10^5 \left(\frac{M_{\rm clus}}{10 {\rm M}_\odot}\right)^{1/2} \left(\frac{R_{\rm clus}}{1000 {\rm AU}}\right)^{3/2} {\rm yr}.
\ee
Therefore, if the stars of the Trapezium Cluster
formed in bound sub-clusters of
$10-100$ stars within $R_{\rm clus} \approx 10^3$ AU, they would have 
evaporated by the present time.  However, sub-clusters with 
$R_{\rm clus} \simgreat 10^4$ should have survived and would be relaxed
(centrally-condensed).  From Figure \ref{orionclus2}, it is possible that
such bound sub-clusters exist in the Trapezium Cluster, but they
cannot be detected by modelling the MSDC.

\begin{figure}
\vspace{0.2truecm}
\centerline{\hspace{0.0truecm}\psfig{figure=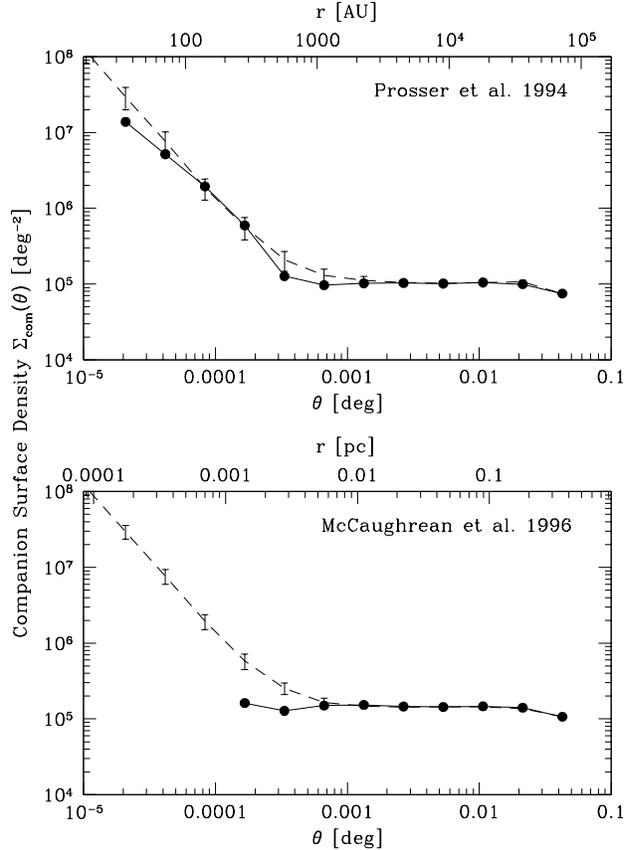,width=11.5truecm,height=11.5truecm,rwidth=9.0truecm,rheight=11.5truecm}}
\caption{\label{trapbinaries} The mean surface density of companions $\Sigma_{\rm com}(\theta)$ for the Orion Trapezium Cluster using the data of Prosser et al. (1994) (top, points and solid-line) and McCaughrean et al. (1996) (bottom, points and solid-line).  The observations are fit by $\Sigma_{\rm com}(\theta)$ of the three-dimensional stellar system distribution given by equation \ref{trapfit}, with 36\% of the systems being binaries with separations in the range $10$ to $10^4$ AU (see text).}
\end{figure}

\subsubsection{Binaries}
\label{orionbinary}

Thus far, we have concentrated on large-scale structure in the Orion 
Trapezium Cluster.  
While the survey of Jones \& Walker \shortcite{JonWal88} does 
not extend to small enough separations to detect binaries above the
confusion limit in the Trapezium Cluster,
the surveys of McCaughrean et al. \shortcite{MRZS96} and, especially, of
Prosser et al. \shortcite{PSHSJWM94} do.  In Figure \ref{trapbinaries} we
model the Trapezium Cluster MSDC, taking into account binaries.  We distribute
stellar systems according to equation \ref{trapfit}, where each stellar
system is randomly chosen to be either a single star or a binary with 
a separation in the range 10 to $10^4$ AU.  Such large maximum binary
separations are possible even though the break between the binary
and large-scale regimes occurs at $\approx 600$ AU since, as seen in
Section \ref{posbreakSFR}, the break does not necessarily give the 
maximum binary separation and in the Trapezium Cluster, 
as seen above, the break is not associated with the Jeans length.
The ratio of binary systems to the total number of systems is 0.36, 
which corresponds to a binary frequency of 60\% if binaries have 
separations between 0.1 and $10^4$ AU that are distributed as in 
equation \ref{eqsepdist}.  For each survey (the Prosser 
et al. survey and the McCaughrean et al. survey) the stars are distributed 
over the same areas as the surveys, and we ensure that the total number of
stars (not systems) is the same as in each survey.  

The fact that the stellar distribution used to model the McCaughrean MSDC 
contains systems that would not have been resolved in the 
survey of McCaughrean et al. does not matter.  What is important is that
the same number of stars are used for the calculation of both MSDCs
(to give the correct normalisation) and that the stars that are in 
close binaries are placed according to the same large-scale spatial 
distribution as the other stars (which, of course, they are).  
An alternative method is to
allocate the same number of resolvable systems in the model distribution
as is observed, and then to multiply the model MSDC by the number of observed 
stars and divide by the number of stars in the model to get the correct
normalisation.

The model gives an excellent fit to {\em both} sets of Trapezium Cluster data 
(Figure \ref{trapbinaries}) which is particularly 
pleasing since equation \ref{trapfit} was derived from fitting only the
McCaughrean MSDC and radial density profiles (not the Prosser et al. data).
The overall binary frequency of 60\% was chosen 
because this agrees with that of the field stars \cite{DuqMay91} 
which in turn is consistent with the binary frequency that Prosser et al.
derived for their data.

The only regions that are not so well fit are the
region between about 0.8 and $2''$ (400 to 1000 AU) and the two smallest 
separation bins of the Prosser MSDC where the Trapezium Cluster has
a lower MSDC than predicted.  The latter of these regions can be 
explained by incompleteness for binaries with close separations.   The
deficit between 400 and 1000 AU, however, gives {\em very weak} evidence 
that binaries with separations $\simgreat 500$ AU may have been 
depleted by binary-single interactions in the centre of the 
Trapezium Cluster
(e.g. Kroupa  1995a,b,c).
A simple calculation of the timescale for such encounters \cite{BinTre87} 
shows that this is possible.  Assuming a central stellar density of
$2 \times 10^4$ pc$^{-3}$
(Section \ref{otc}; Hillenbrand \& Hartmann 1997) 
and a three-dimensional stellar 
velocity dispersion of 4 km s$^{-1}$ 
(Jones \& Walker 1988; Tian et al. 1996), the 
timescale for a star to have an encounter at 500 AU is 
$\approx 7 \times 10^5$ years which is of the same order as the age of 
the Trapezium Cluster \cite{Hillenbrand97}.

\section{Conclusions}
\label{conclusions}

We have studied the interpretation of the mean surface density of 
companions (MSDC) as a function of separation $\Sigma_{\rm com}(\theta)$
in star-forming regions.  We have shown how the power-law slope of 
$\approx -2$ for binaries is due their flat distribution of periods in
the logarithm of separation, and have considered the MSDC of 
various global density profiles, sub-clusters and self-similar distributions.
We emphasise that simply because a power-law 
slope can be fit to a particular MSDC, it does not mean that the stellar 
distribution is self-similar or fractal.
We have also demonstrated the effects of survey
boundaries on the calculation of $\Sigma_{\rm com}(\theta)$.  Several methods
of attempting to avoid boundary effects were considered, all of which 
provide a full correction in the case that there is no large-scale stellar
density gradient across boundaries, but none of which give a perfect 
correction when there are such large-scale gradients.  Of these, we recommend
Method 5, since it allows the maximum range of separations to be studied,
does not discard any information, and is simple to use for surveys with
irregular boundaries.  Even in the case of
a uniform stellar distribution, the improper handing of boundaries
results in the $\Sigma_{\rm com}(\theta)$ having a significant slope
for separations greater than $\approx 1/50$ of the survey area's dimensions
(i.e.~using Method 1).

Larson \shortcite{Larson95} associated the separation at which
a break in $\Sigma_{\rm com}(\theta)$ between the binary regime and 
the large-scale regime occurs with the Jeans length in the Taurus-Auriga
star-forming region (SFR).  However, we show this transition separation 
may only be associated with the Jeans length in special cases, and that the 
transition separation 
does not necessarily give the maximum binary separation.  In general,
the break occurs at the separation where the mean surface density 
of {\em binary} companions is equal to the mean surface 
density of {\em non-binary} companions
(the latter of which may be physically close, or simply chance projections).
Thus, typically, the break occurs at smaller separations for SFRs
with higher stellar surface densities (as observed by Simon 
\shortcite{Simon97} and Nakajima et al. \shortcite{NTHN98}).
In turn, the surface density of non-binary companions 
depends on the parameters of the binaries, 
the volume density of stars, the volume-filling factor of the 
stellar distribution and, in general, the depth of the star-forming region.

The transition separation between the binary and the large-scale 
regimes also evolves with time.  Due to a stellar velocity dispersion,
initial structure is erased and the surface density of stars in an unbound 
region generally decreases.  This effect begins at the smallest scales,
extends to larger scales with time, and results in the transition 
separation increasing with time.  Finally, the transition between the binary
and the large-scale regimes may allow a truncation of binaries at large
separations to be detected, especially in old clusters that
were much denser when the stars were formed and have since expanded.  In such
cases, this provides a record of the stellar density when the stars
first formed.

In summary, the transition separation may be associated with the Jeans
length only if the star-forming region is young enough that initial
structure has not been erased, and if the SFR is `optically thin' in the
sense that projection effects due to the depth of the SFR do not affect the 
transition separation.  The latter is true if the volume-filling factor
of the SFR is low (e.g. the SFR is composed of widely separated clusters
consisting of only a few stars ($\sim 10$), or if the stars have a
fractal distribution with dimension $\simless 1.5$).  This is the case for
the Taurus-Auriga SFR, which explains the good agreement between between the
transition separation and the Jeans length found by Larson 
\shortcite{Larson95}, but it is not the case for the Orion Trapezium Cluster.

It is important when studying the large-scale spatial 
distributions of star-forming regions to obtain the most complete sample
of stars over the largest area possible.  The lack of such data for
the Ophiuchus SFR makes an attempt to study its large-scale spatial
distribution of little use at this time.

For the Taurus-Auriga and Orion Trapezium SFRs, the current data 
makes a meaningful study of their large-scale stellar distribution possible.
For the Taurus-Auriga SFR, Larson \shortcite{Larson95} fit the large-scale
MSDC with a power-law slope that implied a fractal stellar distribution.
However, this is not the only possible interpretation; the data can be 
equally well fit by assuming the stars are formed primarily in 
randomly-distributed clusters of stars.  For the Orion Trapezium SFR,
we find that the MSDC is consistent with the stars simply being distributed
according to a surface density that decreases with radius; there is no
evidence for sub-structure (either fractal or sub-clusters) in the 
stellar distribution.  We also demonstrate how upper limits can be placed on
how much sub-clustering is present, and note the the sensitivity of the
MSDC to detecting sub-structure appears to be slightly less than
that of the human eye.  
The results for the Orion Trapezium SFR are consistent with
the fact that if structure were present when the stars formed,
it would have been erased by the current time due to the
stellar velocity dispersion.

Binaries in the Taurus-Auriga and Orion Trapezium SFR are 
roughly consistent with
an MSDC with a power-law slope of $\approx -2$.  However, we point out that
comparing power-law indices derived from the slope of the MSDC 
in the binary regime is not the
best way to compare the distribution of binary separations between stellar
populations since any structure or deviation from a true power-law may
easily be missed.  In the centre of the Orion Trapezium 
SFR, we find very weak evidence that there may be a deficit of binaries with 
separations $\simgreat 500$\,AU\@.  Such a deficit may be caused by the
disruption of wide binaries by single-binary star encounters.

Finally, in view of our studies of the Taurus-Auriga and Orion Trapezium SFRs,
we emphasise caution when interpreting the MSDC.  Rather than attempting 
to characterise star-forming regions simply by fitting
power-laws to $\Sigma_{\rm com}(\theta)$, it is more instructive to also 
consider the global stellar distribution (e.g. a radial surface density
profile) and to compare the MSDC to those of model stellar distributions to
determine the robustness of any conclusions.  Alternatively, rather than
just considering the MSDC (or, equivalently, the two-point correlation
function), correlation functions of higher order (three and four-point
correlation functions) and/or the nearest-neighbour distribution 
can be used to differentiate between non-hierarchical
and hierarchical structure.  
The use of higher-order correlation functions
is common in studying the large-scale structure of the 
universe \cite{Peebles80}.  The nearest-neighbour distribution has been
used by Nakajima et al. \shortcite{NTHN98} to argue that the power-law
slope of an MSDC on large scales may indicate a stellar age spread rather 
than the presence of hierarchical structure.  However, while an age spread
does help explain their results, we argue that their results do not
exclude the possibility that stars form in hierarchical structures.  
More work is required on this topic.

\section*{Acknowledgments}

We are grateful to Richard Larson, Mike Simon, and the referee, 
Lee Hartmann, for their critical
reading of the manuscript and comments 
which helped improve this paper.  We also thank Michael Meyer, 
Ralf Klessen, Ian Bonnell, and Melvyn Davies for useful discussions.

\end{document}